\documentclass{PoS}
\usepackage[T1]{fontenc} % if needed
\usepackage{textcomp}
\usepackage{slashed}
\usepackage{graphicx}%
\usepackage{subfigure}
\usepackage{mathptmx}
\usepackage{mathrsfs}
\usepackage{bm}% bold math
\usepackage{verbatim}
\usepackage{amsmath}
\usepackage{amssymb}

\newcommand{\beq}{\begin{equation}}
\newcommand{\eeq}{\end{equation}} 
\newcommand{\be}{\begin{eqnarray}}
\newcommand{\ee}{\end{eqnarray}}
\long\def\hidestart#1\hideend{}
\def\be{\begin{eqnarray}}
\def\ee{\end{eqnarray}}

\title{Toward the minimal realization of a light composite Higgs}

\ShortTitle{Toward the minimal realization of a light composite Higgs \hskip 0.9in {\rm Chik Him Wong and}}

\author{Zoltan Fodor\\
        Institute for Theoretical Physics, E\"{o}tv\"{o}s University, H-1117 Budapest, Hungary\\   
        Department of Physics, University of Wuppertal, Gau$\beta$strasse 20, D-42097, Germany\\ 
        J\"{u}lich Supercomputing Center, Forschungszentrum, J\"{u}lich,
        D-52425 J\"{u}lich, Germany}
\author{Kieran Holland\\
        Department of Physics, University of the Pacific, 3601 Pacific Ave, Stockton CA 95211, USA}
\author{\speaker{Julius Kuti}\\
        Department of Physics 0319, University of California, San Diego, 9500 Gilman Drive, La Jolla, CA 92093, USA}  
\author{Santanu Mondal\\
        Institute for Theoretical Physics, E\"{o}tv\"{o}s University,  H-1117 Budapest, Hungary\\
        MTA-ELTE Lendulet Lattice Gauge Theory Research Group, 1117 Budapest, Hungary}
\author{Daniel Nogradi\\
        Institute for Theoretical Physics, E\"{o}tv\"{o}s University, H-1117 Budapest, Hungary\\   
        MTA-ELTE Lendulet Lattice Gauge Theory Research Group, 1117 Budapest, Hungary}
\author{Chik Him Wong*\\
Department of Physics, University of Wuppertal, Gau$\beta$strasse 20, D-42097, Germany\\}
\abstract{\small Work in progress is reported on a particularly interesting gauge theory with a fermion  doublet in the 
two-index symmetric (sextet) representation of the SU(3) color gauge group.
Extending previous studies
we outline our strategy as we 
investigate Goldstone dynamics and  Electroweak scale setting from chiral symmetry breaking (${\rm \chi SB}$), 
test the GMOR relation from the spectrum of the Dirac operator and the related chiral condensate, 
begin to develop and test mixed action based improved analysis of ${\rm \chi SB}$ with new run plans 
at fixed topology to cross over from the p-regime to the epsilon-regime of ${\rm \chi SB}$,
continue to pursue the light ${\rm 0^{++}}$ scalar and its relation to the dilaton,
and probe the scale-dependent running coupling from the
perturbative UV scale to the scale of chiral symmetry breaking.
Our observations suggest that the model is very close to the conformal window and a light composite scalar, perhaps the Higgs impostor
with or without dilaton-like interpretation,
appears to emerge with $0^{++}$ quantum numbers. The lightest baryon of the model on the 3 TeV scale has intriguing implications.}

\FullConference{The 32nd International Symposium on Lattice Field Theory,\\
		23-28 June, 2014\\
		Columbia University New York, NY}

\begin{document}

\section{Introduction}

New physics beyond the Standard Model (BSM) in the framework of some new strongly-interacting gauge
theory with a composite Higgs mechanism is an attractive BSM scenario
with related lattice work reviewed recently~\cite{Kuti:2014epa}.
Of course we hear voices that pursuing the composite Higgs scenario is overtaken by recent findings at the LHC. After all, a light 
Higgs-like scalar was found, consistent with SM predictions, and composite states have not been found below the TeV scale. In contrast,
the voices continue, strongly coupled BSM gauge theories are Higgs-less with resonances predicted below the TeV scale,
just like in the original technicolor idea~\cite{Weinberg:1979bn,Susskind:1978ms}. 
The facts do not seem to support this skeptical view
which originates from naively scaled properties of Quantum Chromodynamics (QCD) to the TeV region.
Related old technicolor guessing games were lacking
any credible predictive power close to the conformal window  where gauge theories are nearly scale invariant, in sharp contrast to QCD which is not.
In fact, there is no evidence that compositeness and a light Higgs scalar are incompatible. 
Recent developments are hinting compatibility, 
like in near-conformal gauge theories where a light composite scalar could perhaps emerge on the Electroweak scale with a 
resonance spectrum far separated above the TeV scale, perhaps within the reach of Run 2 at the LHC. 

Work in progress is reported here on a particularly interesting gauge theory with a fermion  doublet in the 
two-index symmetric (sextet) representation of the SU(3) color gauge group~\cite{hong2004,dietrich2005,plb2012,degrand2012}. 
Our observations suggest that the model is very close to the conformal window and a light composite scalar
appears to emerge with $0^{++}$ quantum numbers.
%perhaps a Higgs impostor with or without dilaton-like interpretation, appears to emerge with $0^{++}$ quantum numbers.
%
From chiral symmetry breaking we find three massless Goldstone pions in the spectrum.
 With Electroweak interactions turned on,  the model exhibits the simplest composite Higgs mechanism 
and leaves open the possibility
of a light composite scalar state with quantum numbers of the Higgs impostor emerging as perhaps the 
pseudo-Goldstone dilaton-like state from spontaneous symmetry breaking of scale invariance. 
Even if scale symmetry breaking is entangled with
$\chi{\rm SB}$  without dilaton interpretation, 
a light Higgs-like scalar state can emerge  from the new gauge force close to the conformal window.
The main goal of our  Higgs project is to investigate these important problems 
with ab initio lattice simulations of the sextet model.

In Section 2 we introduce the Electroweak embedding of the strongly coupled sextet gauge theory, comment
on the intriguing features of the lowest baryon state in the minimal sextet model and its extensions, and describe 
the new data sets developed since our last report~\cite{Fodor:2014pqa,Fodor:2014zca}. 
In section 3 we investigate Goldstone dynamics and Electroweak scale setting 
from chiral symmetry breaking as premier ingredients of the composite Higgs mechanism.
We also analyze cutoff dependent taste breaking effects in the non-Goldstone pion-like spectrum of staggered fermions.
In Section 4 we test the GMOR relation from the spectrum of the Dirac operator and the related chiral condensate. 
In Section 5 we present new results on the  light ${\rm 0^{++}}$ scalar and outline future plans.
In Section 6 we begin to develop and test mixed action based improved analysis of ${\rm \chi SB}$ with new run plans 
at fixed topology to cross over from the p-regime to the epsilon-regime of ${\rm \chi SB}$.
In Section 7 we probe the scale-dependent running coupling from the
perturbative UV scale to the scale of chiral symmetry breaking.

%\newpage
\section{Electroweak embedding and computational framework}

\noindent{\bf Quantum numbers and Electroweak symmetry breaking pattern}

The two fermion flavors of the model transform in the complex two-index symmetric (sextet) representation of the SU(3) color gauge group
which implies ${\rm SU(2)_L\times SU(2)_R\times U(1})$  flavor symmetry for the gauge force of the theory.
The fermions  are assembled into a left-handed weak isospin doublet ${\rm q_L}$ and two right-handed weak isospin singlets  ${\rm q_R}$ with
%\begin{equation}
%
\begin{align}
  q^{(i,j)}_L =
  \begin{bmatrix}
    u^{(i,j)}_L \\ d^{(i,j)}_L 
  \end{bmatrix}, \qquad\qquad
&
  q^{(i,j)}_R =
  \begin{bmatrix}
    u^{(i,j)}_R,  & d^{(i,j)}_R
  \end{bmatrix},
 \label{eq:1}
\end{align}
where ${\rm  i,j=1,2,3}$ label the color indices of the symmetric tensor elements. 
The ABJ anomaly would spoil the renormalizability of the gauge theory so 
the fermion gauge coupling must not introduce anomalous Ward indentities. This requires
the trace ${\rm tr(\{T^a(R),T^b(R)\}T^c(R))}$ to vanish for fermion flavor group
representation R with representation matrix ${\rm T^a(R)}$.
In the sextet model, the fermions are either doublets or singlets under the flavor group SU(2). 
The matrix ${\rm T^a}$ will be either the Pauli matrix ${\rm \tau^a}$
or the U(1) hypercharge Y. Since the SU(2) group is anomaly-free, 
${\rm tr(\{\tau^i,\tau^j\}\tau^k)=2\delta^{ij}tr(\tau^k)=0}$, it is easy to see that ${\rm tr (Y)=tr(Y^3)=0}$ 
are the two anomaly condition to satisfy.
The absence of gauge anomalies thus requires 
a traceless weak hypercharge operator 
for the left-handed fermion doublet. Choosing the generator of the U(1) group as  ${\rm Y=2(Q-T_3)}$ for 
the weak hypercharge Y, with charge Q and the
third component ${\rm T_3}$  of weak isospin, requires half-unit of charge for the left-handed fermions which form a weak isospin doublet, 
\be
  q^{(Q)}_L =
  \begin{bmatrix}
    u^{(1/2)}_L \\ d^{(-1/2)}_L 
  \end{bmatrix}.
  \label{eq:2}
  \ee
The right-handed u-fermion will have hypercharge ${\rm Y=1}$  and the right-handed d-fermion will have hypercharge ${\rm Y=-1}$ 
with consistent charge assignments
\be
  q^{(Q)}_R =
  \begin{bmatrix}
    u^{(1/2)}_R, & d^{(-1/2)}_R 
  \end{bmatrix},
 \label{eq:3}
\ee
satisfying the ${\rm tr(Y^3)=0}$ anomaly condition.
The chiral ${\rm SU(2)_L\times SU(2)_R}$ symmetry of the theory is dynamically broken to the diagonal vector ${\rm SU(2)_V}$ 
subgroup. This ${\rm \chi SB}$  is responsible for breaking ${\rm SU(2)_W\times U(1)_Y}$ to ${\rm U(1)_{em}}$.
The residual ${\rm SU(2)_V}$ symmetry is the electroweak analog of isospin and the approximate weak isospin invariance 
of the electroweak force and the new gauge force ensure that the ${\rm \rho}$-parameter is approximately one.
There is also an exact U(1) symmetry in the theory which is unaffected by chiral symmetry breaking protecting the conserved baryon number. 
Baryon number and charge conservation are keeping the lightest baryon stable against the new gauge force and weak decays. 

\vskip 0.1in
\noindent{\bf Baryon construction in the sextet model}

The charge assignment has intriguing implications for the baryon spectrum of the sextet model. The sextet representation of fermions with 
SU(3) color gauge group will impose a symmetric color wave function for  baryon states as three-fermion systems. This is in sharpe contrast
to QCD where the color wave function of the nucleon is antisymmetric. The non-relativistic limit of the flavor-spin-spatial part of the baryon wave function will
look like the wave function of triton in terms of symmetries~\cite{Blatt:1958}. It follows from the construction of Eqs.~(\ref{eq:1}-\ref{eq:3})
that the lightest baryons form a stable isospin doublet of (uud) and (udd) states which carry half-integer charges with opposite sign. 
Detailed properties of the sextet baryons with lowest mass in the 3 TeV range were reported at the conference~\cite{Santanu:2015}.
Dark matter related relic sextet baryon issues are expected to come into focus only if the model will deliver a viable composite Higgs
mechanism which remains the primary focus of our investigations.

\vskip 0.1in
\noindent{\bf Dark matter}  

The lowest stable baryon state in the sextet model is in the fractionally 
charged massive particle (FCHAMP) category of dark matter speculations regarding the 
evolution history of the Universe~\cite{DeRujula:1989fe,Langacker:2011db}. 
It is difficult and ultimately imperative to estimate the relic abundance of the 
fractionally charged and stable sextet baryon. Only qualitative arguments can be given that the very small relic abundance 
is likely to escape existing experimental limits and theoretical requiem~\cite{Langacker:2011db}.
Sextet baryons and antibaryons are produced in the Electroweak phase transition which is expected to be of second order
with two fermion flavors in the chiral limit. Charge symmetric baryon and antibaryon densities in thermal equilibrium will
continue to decrease well below the critical temperature ${\rm T_c}$ until at freeze-out temperature ${\rm T_*}$ 
the expansion rate will set the 
density to its relic abundance level from the solution of the Boltzman equation~\cite{Steigman:1979kw}.
The freeze-out temperature  and the related relic abundance level are very sensitive to the annihilation rate of baryons and
antibaryons in the sextet model. As a general trend, the stronger the interaction, the longer the particles 
remain in equilibrium (larger ${\rm x_*=M_b/T_*)}$ and the fewer survive (${\rm \sim e^{-x_*}}$ ).
Earlier technicolor estimates which were based on scaled up QCD calculations of annihilation cross sections are
irrelevant for the new theory which is close to the conformal window. 
The expected relic abundance is minuscule but more quantitative calculations are required with some sense of urgency
to settle this interesting question.

Some surviving sextet baryon asymmetry in the early evolution of the Universe could affect the above argument  based on a symmetric
sextet baryon-antibaryon distribution. B-violation as required by the first of the Sakharov conditions can be associated with
the non-Abelian anomaly of the left-handed fermion current in nontrivial background gauge field configurations
outside perturbation theory ~\cite{Cline:2006ts}. For the second Sakharov condition,
the origin of C and CP violation, if any, is left undetermined in the new theory and the well-known effects related to sphaleron
dynamics could wash out any early sextet baryon asymmetry generated before the Electroweak phase transition~\cite{Cline:2006ts}. 
%As argued below, the abundance of relic sextet baryons from the evolution history of a charge symmetric  Universe is expected to remain far below 
%detectable levels. 
Explaining dark matter would probably require an extension of the minimal sextet model by adding 
a new lepton doublet with quantum number assignments of the doublets and singlets
following the QCD pattern. 
It could make the lowest baryon state neutral but still requires efficient C and CP violation of ill-understood origin. 
As an alternate extension, a third fermion flavor could be added which is massive and remains an Electroweak singlet.

\vskip 0.1in
\noindent{\bf Algorithms, codes, and run parameter sets}

We use the tree-level Symanzik-improved gauge action for all simulations reported here.
The conventional $\beta=6/g_0^2$ lattice gauge coupling is defined as the overall
factor in front of the well-known terms of the Symanzik lattice action.  
%We have a new data set at  $\beta=3.15,3.20,3.25,3.30$ from our simulations.
The link variables in the staggered fermion matrix are exponentially smeared with  two
stout steps~\cite{Morningstar:2003gk}; the precise definition of the staggered stout action was given earlier in~\cite{Aoki:2005vt}  
and the RHMC algorithm has been deployed in all runs. The fermion flavor doublet requires rooting in the algorithm.
For molecular dynamics time evolution we apply multiple time scales~\cite{Urbach:2005ji} and the
Omelyan integrator~\cite{Takaishi:2005tz}. We have highly efficient codes running on BG/Q, gpu, 
and commodity cluster platforms. 
Our error analysis of  hadron masses is based on correlated fitting with double jackknife 
procedure on the covariance matrices~\cite{DelDebbio:2007pz}.
The time histories of the fermion condensate, the gauge field energy on the gradient flow, the topological charge, 
and correlators are all used to monitor autocorrelation times in the simulations.
We have new simulation results at $\beta=3.2$ and $3.25$ for fermion masses 
${\rm m=0.002, 0.003, 0.004}$ on
$32^3\times64$,  $40^3\times80$,  and $48^3\times96$ lattice volumes.  
We also have new runs at $\beta=3.15$ with ${\rm m=0.003, 0.004, 0.006, 0.008}$ 
and at $\beta=3.30$ with ${\rm m=0.005, 0.006, 0.008, 0.010}$ on
$32^3\times64$ lattice volumes.

%\vskip -0.5in
%\newpage 
%\section{Taste breaking analysis of the non-Goldstone pion spectrum}
\section{ Goldstone spectrum and Electroweak scale setting}
If the chiral ${\rm SU(2)_L\times SU(2)_R}$ symmetry of the model is dynamically broken to the diagonal vector ${\rm SU(2)_V}$
three associated Goldstone pions facilitate
the minimal realization of the Higgs mechanism after the Electroweak interactions are turned on.
\begin{figure}[htb!]
\begin{center}
\begin{tabular}{cc}
\includegraphics[height=5cm]{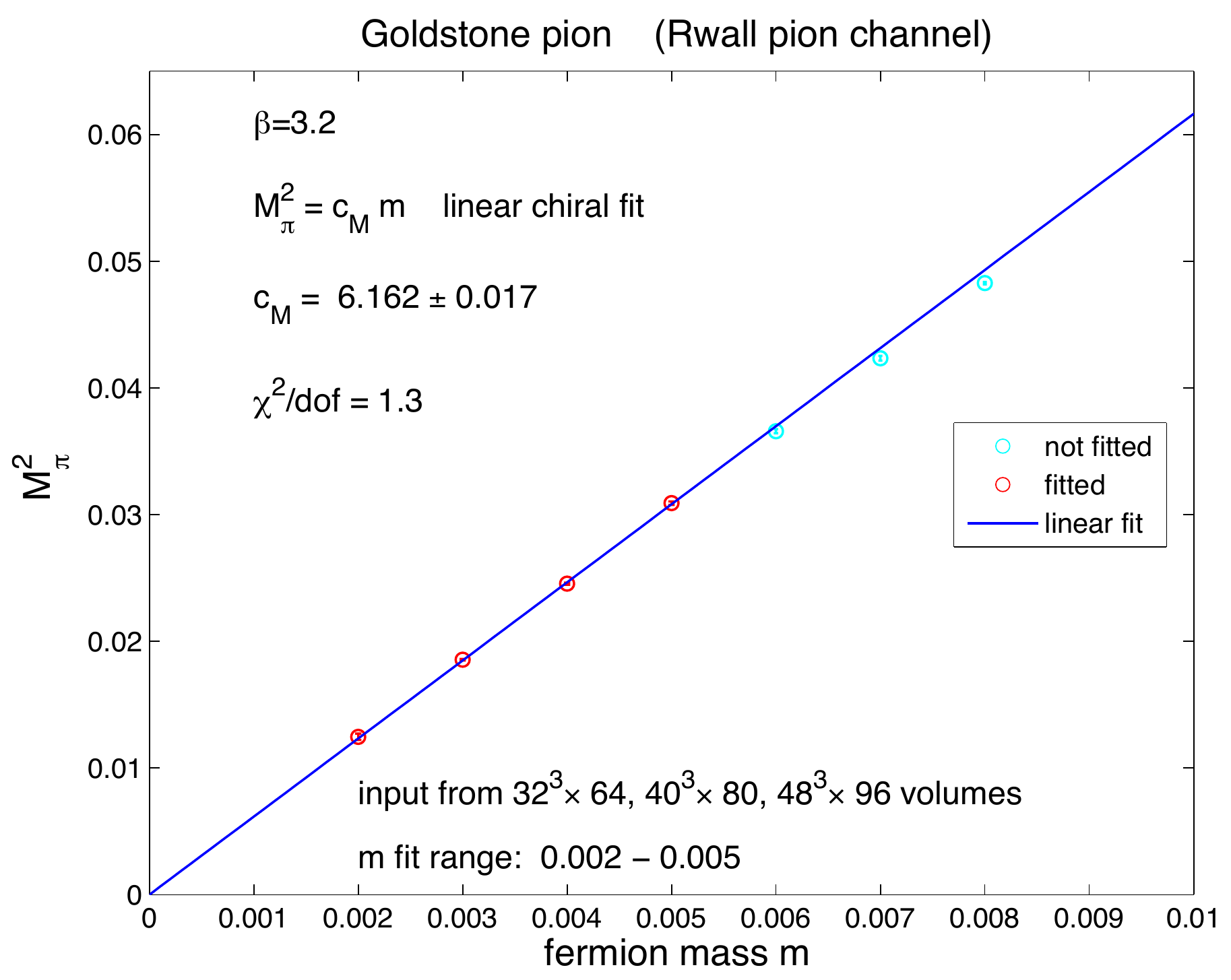}&
\includegraphics[height=5cm]{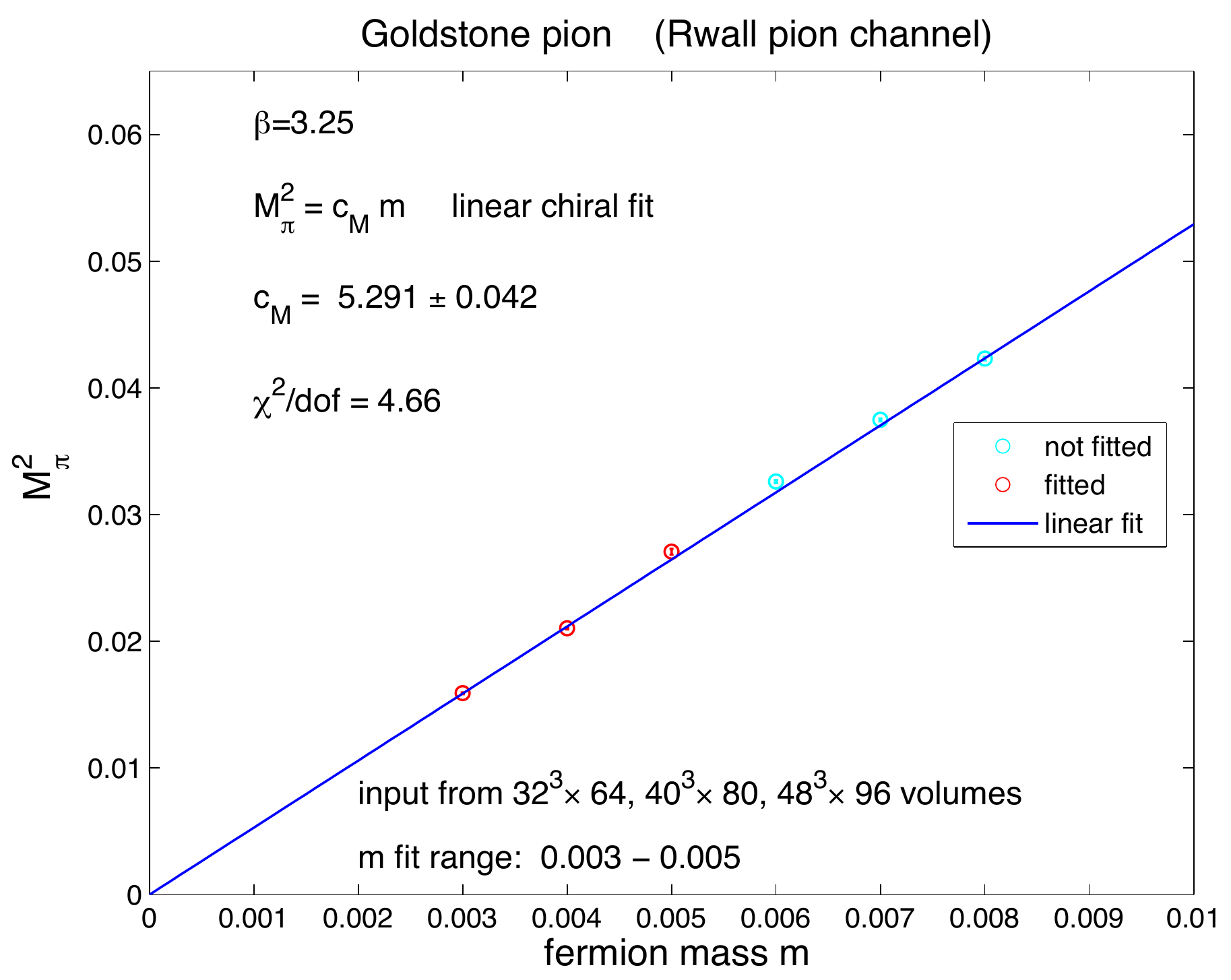}\\
\includegraphics[height=5cm]{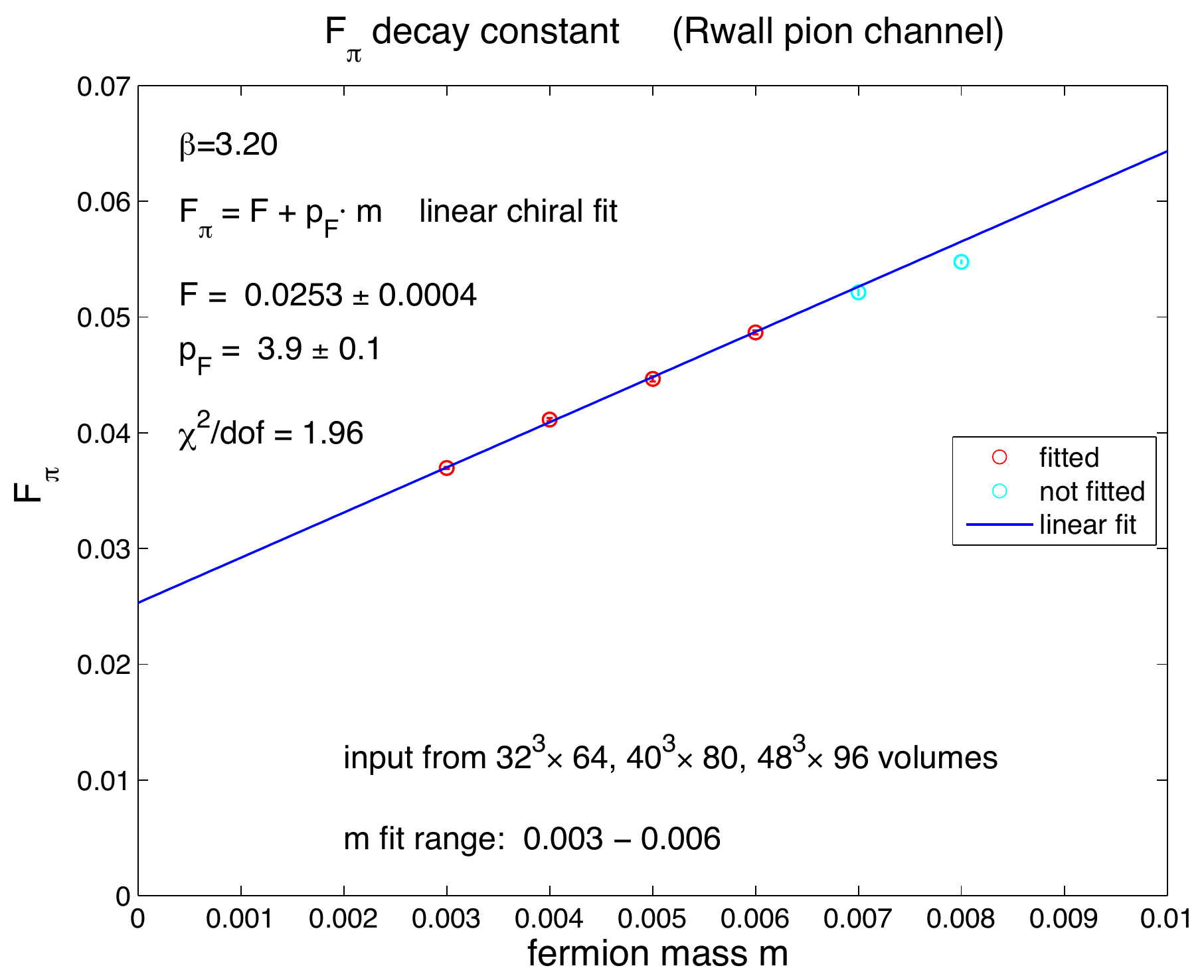}&
\includegraphics[height=5cm]{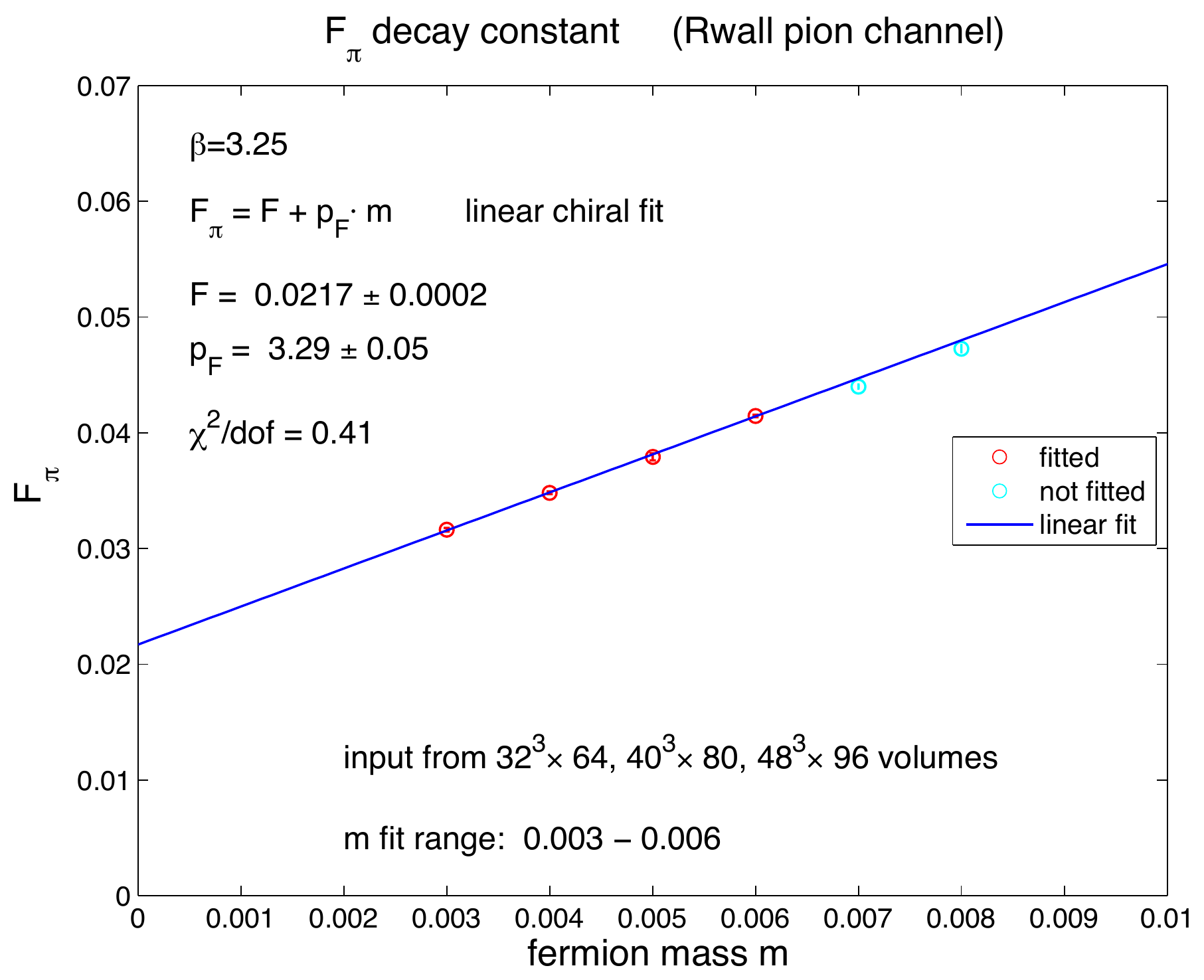}
\end{tabular}
\end{center}
\vskip -0.2in
\caption{\footnotesize  The leading order fits are shown at two values of ${\rm \beta=6/g^2_0}$ 
for the fermion mass dependence of the Goldstone pion from chiral perturbation theory ${\rm (\chi PT)}$ 
without logarithmic loop corrections. Ignoring taste breaking the fundamental ${\rm B}$ parameter of 
the chiral Lagrangian is given by ${\rm C_M/2}$ in leading order (LO).
The fits of  $F_\pi$ include the linear NLO analytic term from  ${\rm \chi PT}$.
All fits are based on the staggered pion correlator with
exact PCAC relation using random wall noise vectors. 
We have similar analysis for ${\rm M^2_{\pi}}$ and ${\rm F_{\pi}}$ at ${\rm \beta=3.15~and~3.30}$ as well. }
\label{fig:PionSpectrum}
%\vskip -0.2in
\end{figure}
As shown in Figure~\ref{fig:PionSpectrum}, our results are consistent with chiral symmetry breaking 
exhibiting consistent Goldstone pion behavior under fermion mass deformations. The Electroweak scale in finite 
lattice spacing units is set from the pion decay constant ${\rm F_\pi}$ in the chiral limit with ${\rm F=250~GeV}$ in continuum physics
notation. The preliminary results of  Figure~\ref{fig:PionSpectrum} represent work in progress with moving parts which include 
continuing refinement of the fitting procedures on our large new data set, the unfinished analysis of taste breaking in staggered 
chiral perturbation theory, new runs closer to the p-regime of leading chiral logarithms, and the influence of a light scalar state 
on the analysis of chiral perturbation theory. In Section 6 we will briefly sketch new directions 
%which include ongoing run sets 
with crossover from the p-regime to the epsilon regime and Random Matrix Theory (RMT) applying mixed actions in the analysis. 
\vskip 0.1in
\noindent{\bf Taste breaking cutoff effects}

Since the determination of the Goldstone decay constant ${\rm F}$ in the chiral limit is critically important for the determination of the 
light scalar mass and the separated resonance spectrum, we will briefly describe taste breaking effects which will influence the final outcome
of the analysis.
\begin{figure}[htb!]
\begin{center}
\begin{tabular}{cc}
\includegraphics[height=5cm]{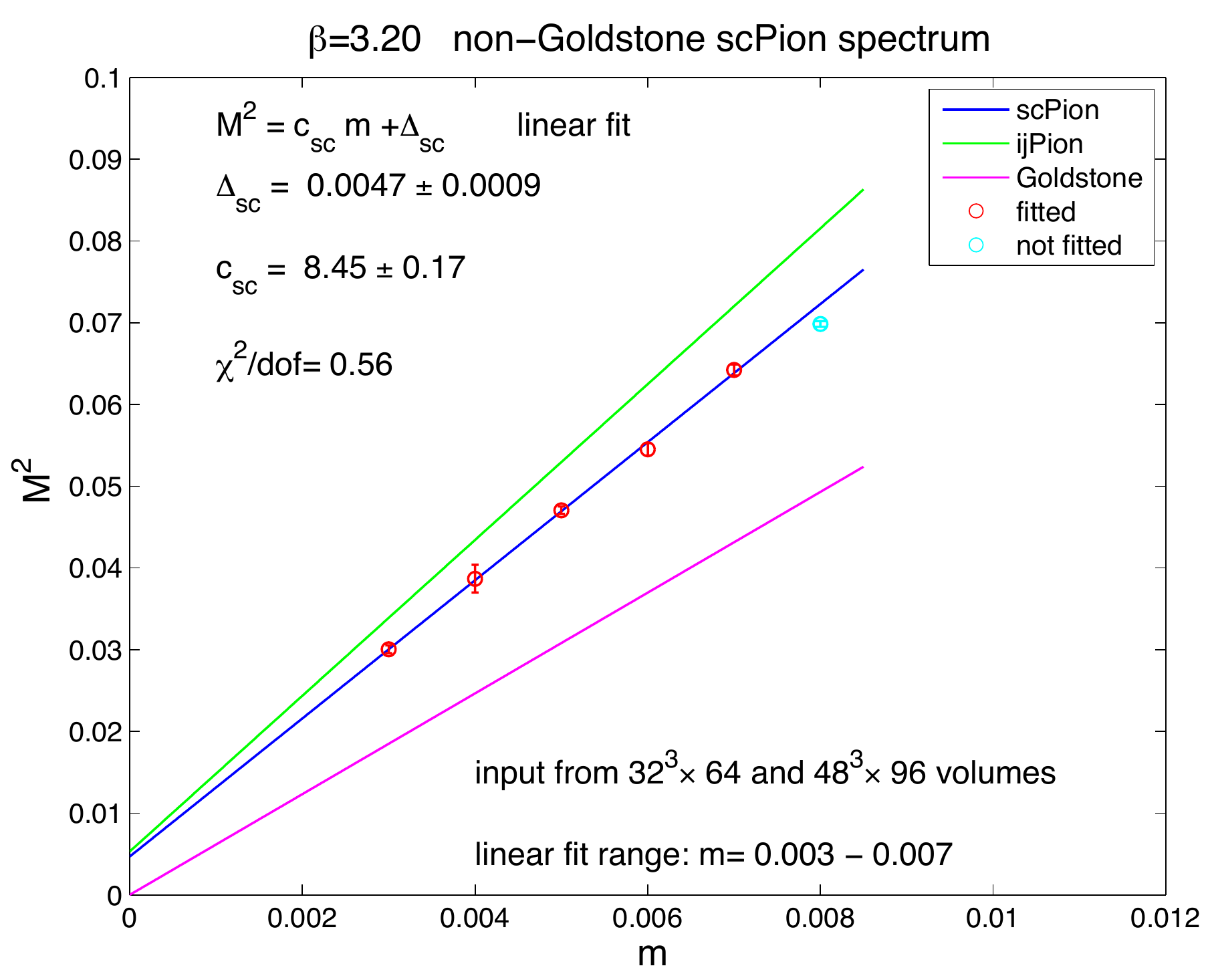}&
\includegraphics[height=5cm]{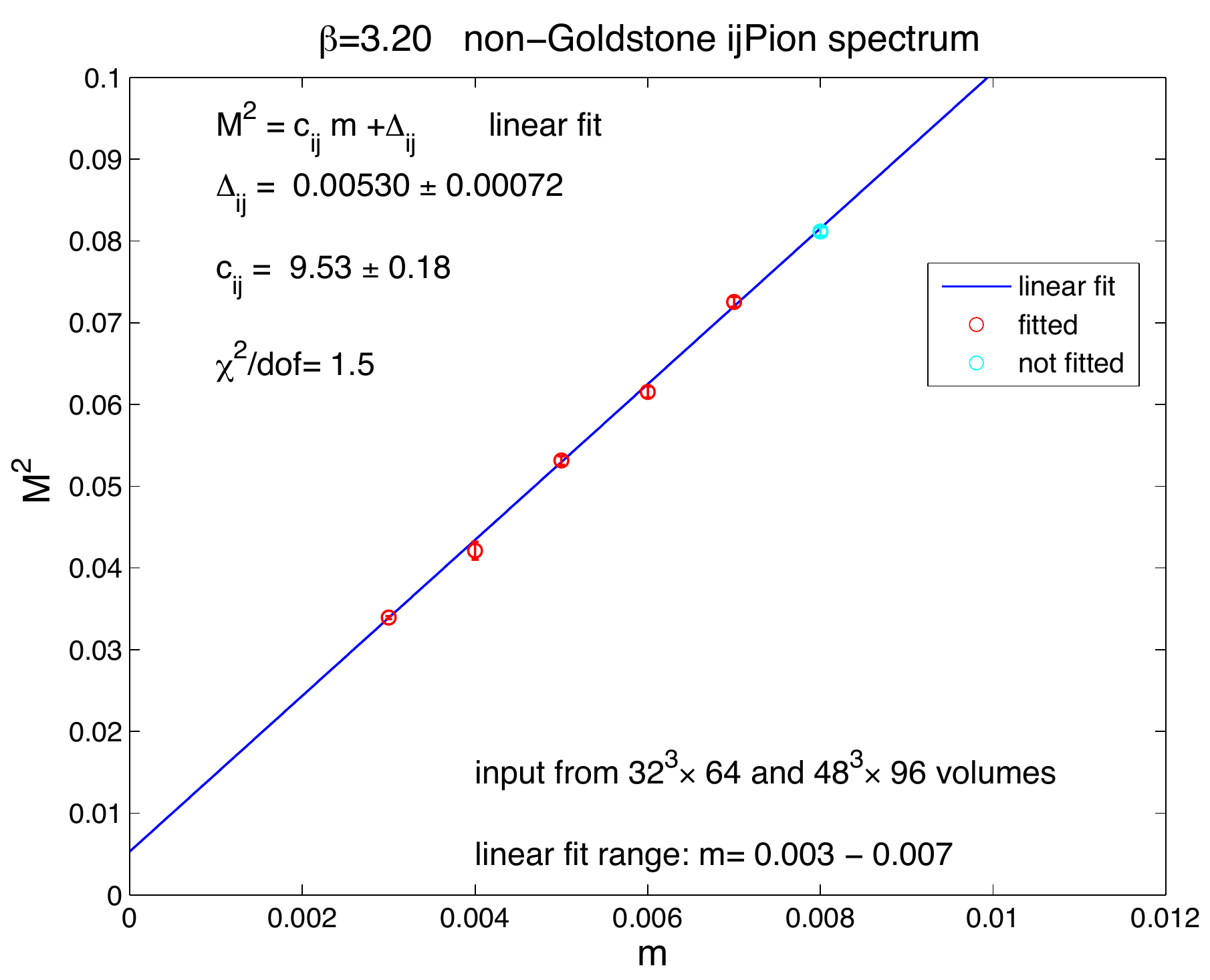}\\
\includegraphics[height=5cm]{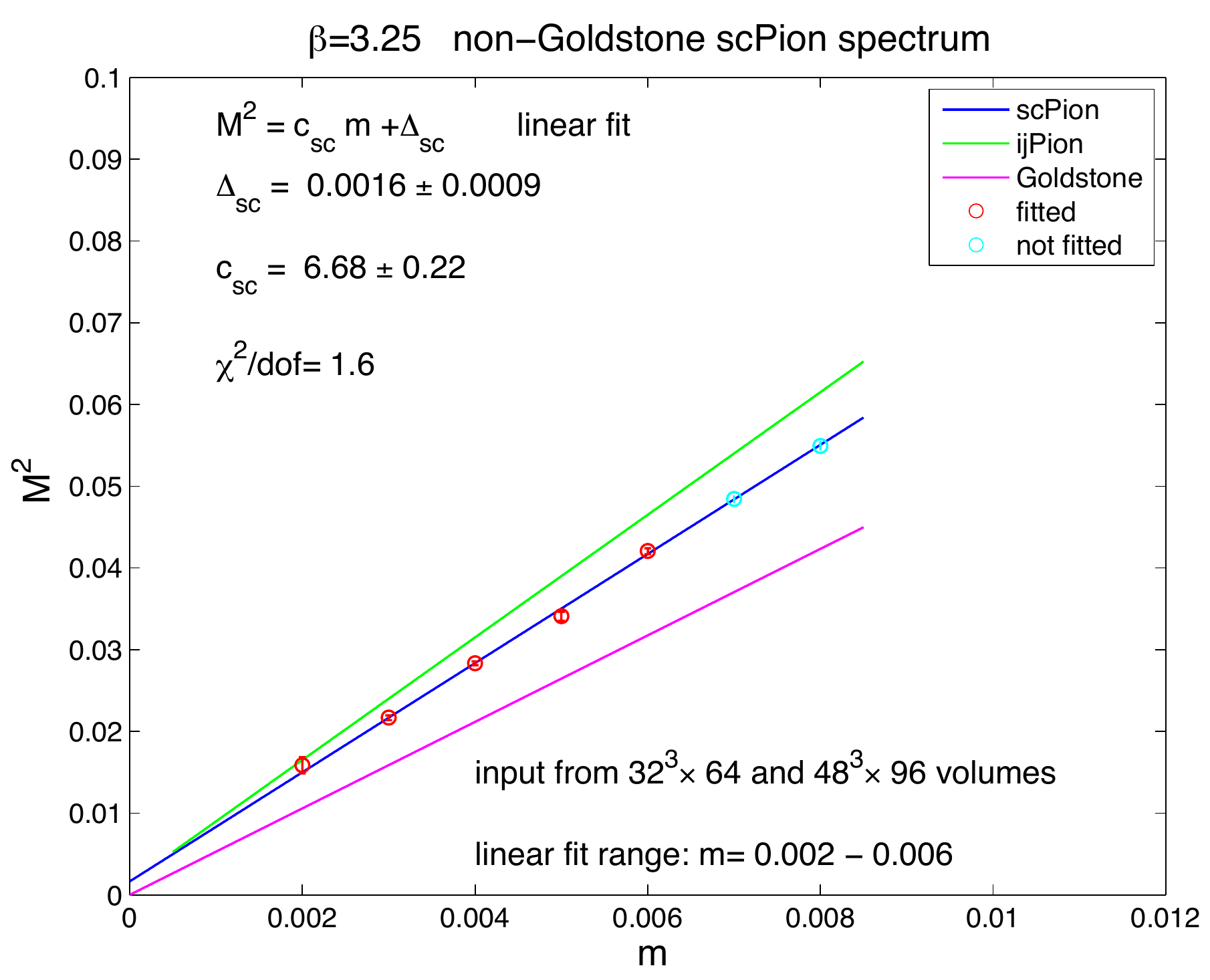}&
\includegraphics[height=5cm]{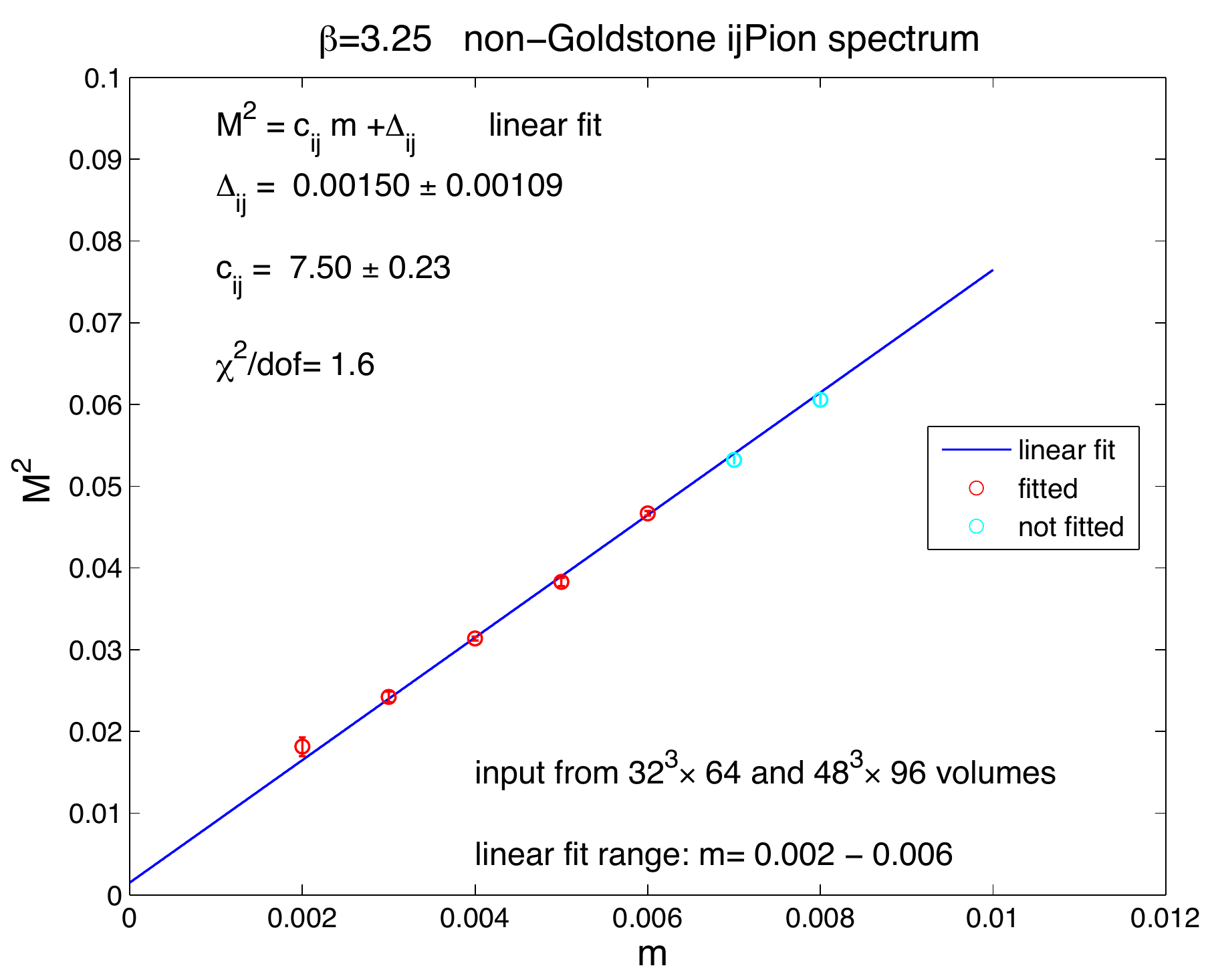}
\end{tabular}
\end{center}
\vskip -0.2in
\caption{\footnotesize  LO fits from the analytic mass dependence of the chiral Lagrangian without logarithmic 
chiral loop corrections are shown for two non-Goldstone pions  at two values of ${\rm \beta=6/g^2_0}$.
We have similar fits at ${\rm \beta=3.30}$. The scPion is degenerate with the i5Pion (not shown)  
in the same SO(4) multiplet~\cite{Lee:1999zxa}.}
\label{fig:non-Goldstone}
%\vskip -0.2in
\end{figure}
To illustrate cutoff dependent taste breaking effects, spectra of selected non-Goldstone pion states are analyzed
in Figure~\ref{fig:non-Goldstone} with the definition of the relevant correlators and
quantum numbers given in~\cite{Fodor:2011tu,Fodor:2012ty}. In the fermion mass range of our data set the taste breaking pattern 
is different from QCD where the residual ${\rm \Delta}$ mass shifts of the non-Goldstome pions are equispaced
in the chiral limit with approximately degenerate SO(4) taste multiplets and with  parallel slopes for finite fermion mass deformations 
of Goldstone and non-Goldstone pion states~\cite{Lee:1999zxa}.
%In the sextet model the residual mass shifts  of the non-Goldstone pions are not equispaced.
For example, as part of the equispaced split of degenerate SO(4) multiplets in QCD, the observed approximate split
${\rm \Delta_{ij} \sim  2\Delta_{sc}}$ of two multiplets appears to have collapsed in the sextet model from our fitting procedure.

The other distinct difference from QCD is the non-parallel slopes which fan out  
in Goldstone and non-Goldstone  mass deformations of the pion spectrum as shown
in Figure~\ref{fig:non-Goldstone}. While the ${\rm \Delta}$ additive mass shifts are LO taste breaking effects 
in the chiral Lagrangian~\cite{Lee:1999zxa,Aubin:2003mg}, the taste breaking slope corrections ${\rm \delta}$
can plausibly be identified with NLO analytic terms in the chiral analysis~\cite{Sharpe:2004is}. The corrected mass 
relation is ${\rm M^2_{NLO} = M^2_{LO}(1+\delta)}$
where ${\rm \delta}$ depends on the taste quantum number of the pion state. 
Several relations
constrain the ${\rm \delta}$ taste breaking corrections~\cite{Sharpe:2004is}.
For example the relation ${\rm \delta_{\pi} = -\delta_{ij}}$ immediately implies that the fitted slope of the Goldstone pion 
must receive significant taste breaking and
cutoff dependent correction in the linear fit of Figure~\ref{fig:PionSpectrum} since the slopes of the Goldstone pion and ijPion fan out 
considerably in Figure~\ref{fig:non-Goldstone}. We can infer from the measured slopes the important
relation  ${\rm \delta_{\pi} = - (c_{ij} - c_M)/(c_{ij} + c_M)}$ to determine the leading cutoff correction to
the fundamental ${\rm B}$ parameter of the chiral Lagrangian from the fit parameter ${\rm c_M}$ as 
${\rm 2B=c_M/(1+\delta_{\pi})}$.
The correction factor ${\rm \delta_{\pi}}$ decreases from -0.207 to -0.087 as ${\rm \beta}$ is varied 
from 3.20 to 3.30 with decreasing lattice spacing.
Work on cutoff corrections to the decay constant ${\rm F}$ are in progress.
The small ${\rm \Delta}$ mass shifts in the chiral limit and the significant fan-out taste breaking structure of the slopes led us to generate
a new data set below the 
fermion mass region ${\rm m=0.002-0.006}$. 
This new effort crossing over from the p-regime toward the epsilon regime and  RMT using mixed action based analysis 
will be outlined in Section 6.

%\newpage
\section{Chiral condensate and GMOR}
The consistency of the fundamental parameters ${\rm F ~ and~B}$ and the direct
\begin{figure}[h!]
\begin{center}
\begin{tabular}{cc}
\includegraphics[height=5cm]{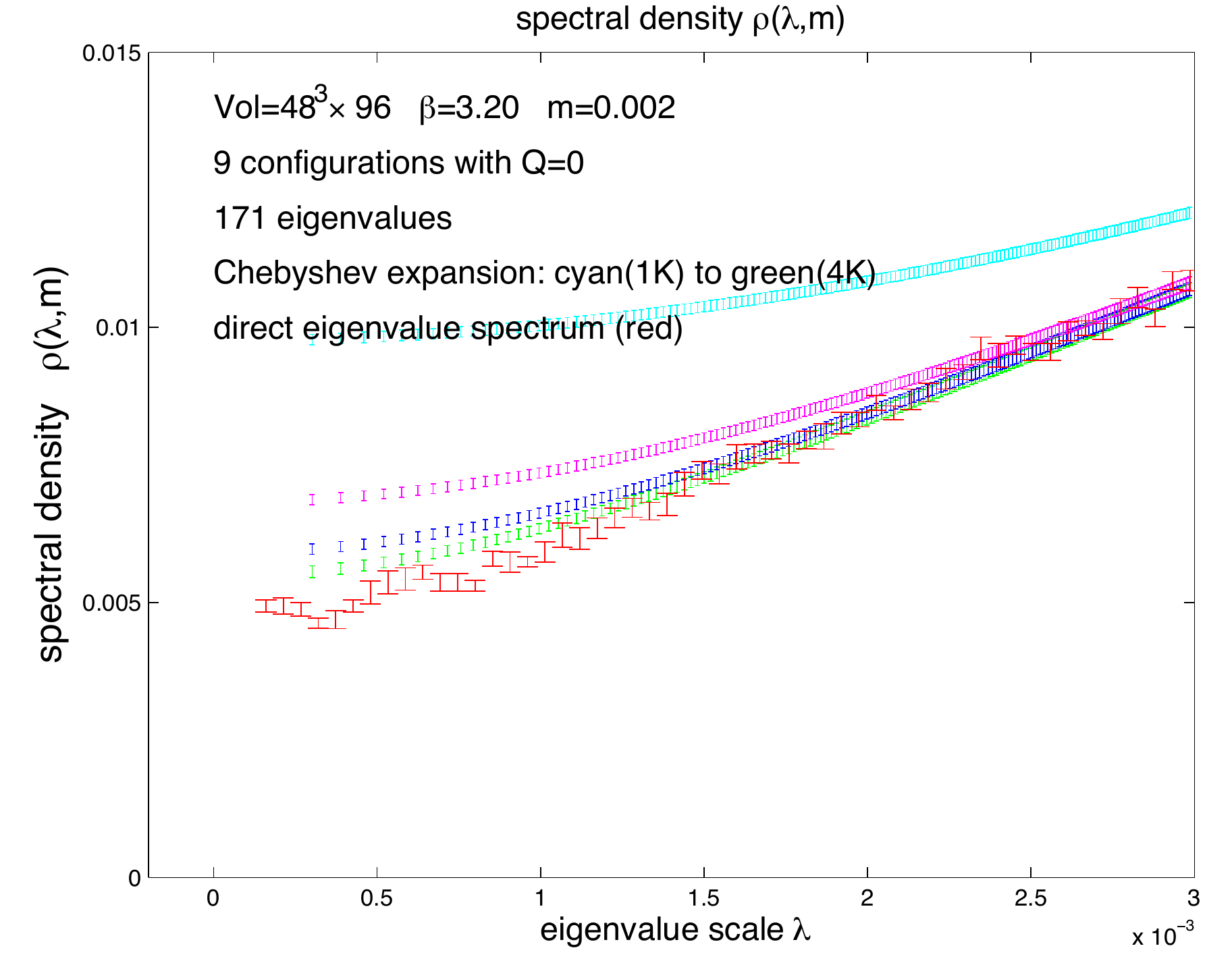}&
\includegraphics[height=5cm]{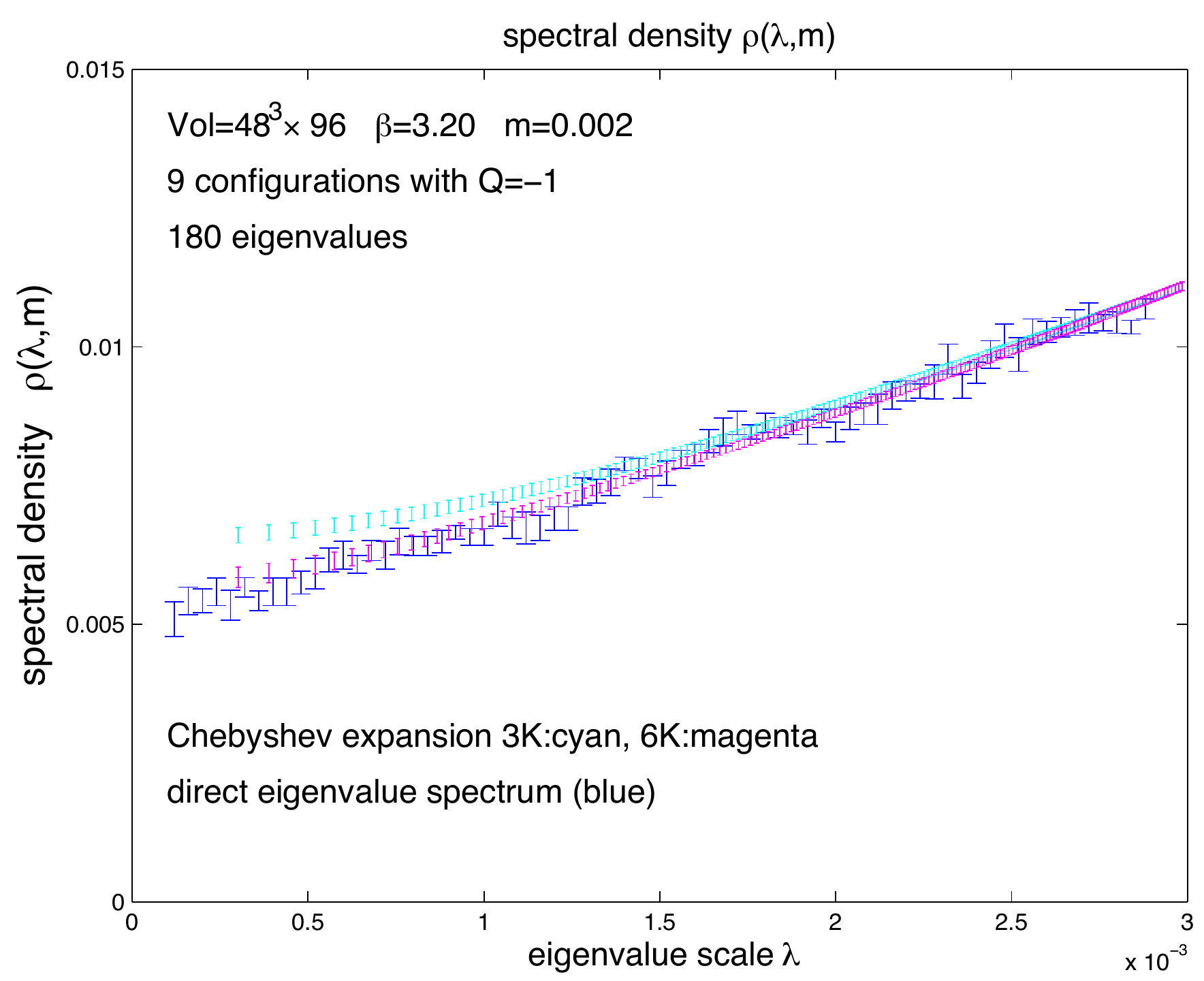}\\
\includegraphics[height=5cm]{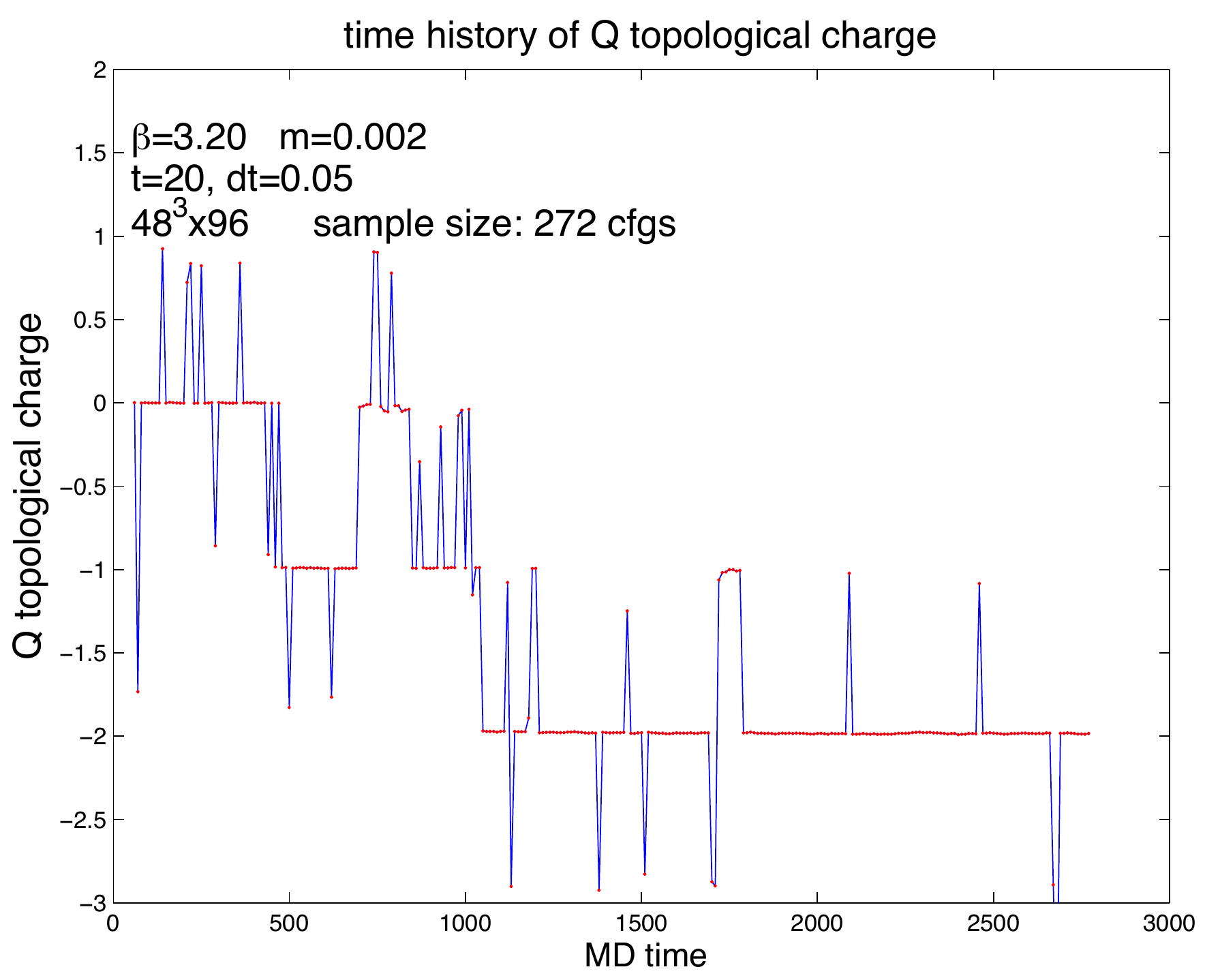}&
\includegraphics[height=5cm]{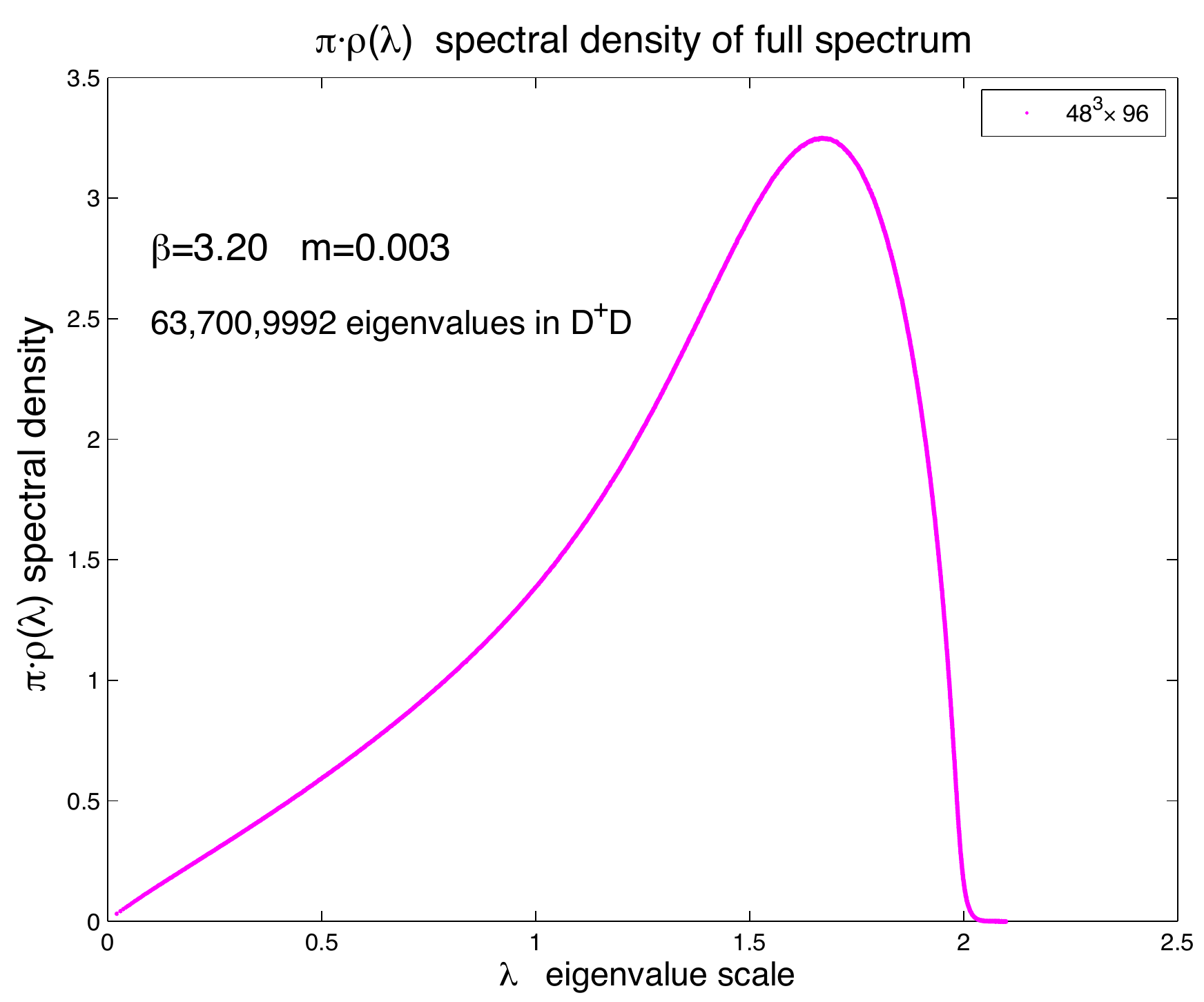}
\end{tabular}
\end{center}
\vskip -0.2in
\caption{\footnotesize Tests on the spectral density of the staggered Dirac operator are performed on our largest ${\rm 48^3\times 96}$ 
lattice volumes at ${\rm \beta=3.20}$ with two different fermion masses. Our new Chebyshev expansion is also shown.}
\label{fig:condensate}
%\vskip -0.2in
\end{figure}
determination of the non-vanishing fermion condensate  ${\rm \Sigma}$ in the chiral limit is tested by the
GMOR relation ${\rm 2BF^2 = \Sigma}$ where summation over two fermion flavors is included~\cite{GellMann:1968rz}. 
Access at ${\rm \beta=3.20}$ to the numerical estimate ${\rm 2BF^2} = 0.00497$ within the range of a few percent error is provided 
by the analysis of data and fits shown in Figure~\ref{fig:PionSpectrum} and Figure~\ref{fig:non-Goldstone}.
The slope correction ${\rm \delta_{\pi}}$ is a significant factor in the analysis. 
The Banks-Casher relation~\cite{Banks:1979yr} relates the condensate ${\rm \Sigma}$ to the spectral density 
${\rm \rho(\lambda,m)}$ of the Dirac operator,
\be
\rho(\lambda,m) = \sum_{k=1}^{\infty} \langle \delta(\lambda - \lambda_k) \rangle/V, \quad
{\rm with}\quad
%The Banks-Casher relation for two fermion flavors is given by 
\lim_{\lambda \rightarrow 0} \lim_{m \rightarrow 0} \lim_{V \rightarrow \infty} \rho(\lambda,m) = \Sigma/(2\pi),
\label{eq:Banks}
\ee
where the spectral density is determined as the ensemble average over the Dirac eigenvalue density in finite volume ${\rm V}$. 
In Eq.~(\ref{eq:Banks}) the condensate  $\Sigma = - \langle \bar{\psi} \psi \rangle$ for two fermion flavors  is determined by the eigenvalue density.
Figure~\ref{fig:condensate} shows a select subset of recent analysis of the spectral density 
of runs with ${\rm 48^3\times 96}$ lattice volume at ${\rm \beta=3.20}$ for the two lowest fermion masses ${\rm m=0.002,0.003}$. 
The lower left panel is the topological history of the ${\rm m=0.002}$ run on the gradient flow at flow time ${\rm t=20}$.  The upper left panel 
shows the spectral density (with the ${\rm 2\pi}$ factor absorbed) for a select subset of gauge configurations with topological charge ${\rm Q=0}$ and
the upper right panel with ${\rm Q=-1}$. The results on the spectral density  ${\rm \rho (\lambda=0,m)}$, at the lowest values of ${\rm \lambda}$ 
reached for the ${\rm \lambda\rightarrow 0}$ limit, are not sensitive to the two values of the topological charge tested. Agreement with GMOR is remarkably 
good from the independent few percent level determination of ${\rm 2BF^2} = 0.00497$ as discussed above. The new analysis removes earlier
inconsistencies from the sextet GMOR relation~\cite{Fodor:2012ty}. 
Continued work is necessary for a more complete analysis of the systematic effects. Comprehensive finite size scaling analysis,
the chiral limit ${\rm m\rightarrow 0}$ of the spectral density, and the scale-dependent renormalization of the condensate remain 
important unfinished goals.

The determination of the spectral density from the low eigenvalues  of the Dirac operator has a limited range
and becomes increasingly difficult for larger volumes.  In several applications, like the anomalous dimension of the mode number density,
 it is important to determine ${\rm \rho(\lambda,m)}$ 
for a large range of ${\rm \lambda}$ and in large lattice volumes.
Recently we developed and tested a new stochastic method with random noise vectors
which is capable of calculating the entire spectral density function and 
mode number distribution of the Dirac operator with great efficiency~\cite{kuti:condensate}. 
The method is based on a high precision finite resolution Chebyshev approximation 
to the Dirac delta function in Eq.~(\ref{eq:Banks}). At any given value of ${\rm \lambda}$  in the spectrum
we can extrapolate to infinite polynomial order in the Chebyshev expansion utilizing well-known asymptotic 
properties of  ${\rm T_n(\lambda)}$ Chebyshev polynomials 
at fixed ${\rm \lambda}$ in the ${\rm n\rightarrow \infty}$ limit. The expansion used here is different from 
the mode number approximation introduced earlier at a fixed value of ${\rm \lambda}$~\cite{Giusti:2008vb}. 
%In the Chebyshev expansion of the Dirac delta function the spectral
%density and mode number density are determined over the full range of  ${\rm \lambda}$ using stochastic evaluation 
%based on random noise vectors.

The lower right panel of Figure~\ref{fig:condensate} 
displays a typical result on our largest sextet lattice volume, averaged over gauge configurations for the full 
eigenvalue spectrum. When magnifying the low infrared part of the spectral density with the same expansion,
the two upper panels show the convergence rate of the Chebyshev approximation as a function of the the polynomial order
when compared with the direct diagonalization of the Dirac operator. Polynomial order in the ${\rm n=6000}$ range is almost 
indistinguishable from the data and the extrapolation procedure works well from lower orders.
There is a variety of interesting applications where this method can be further explored.

\section{The ${\mathbf 0^{++}}$ light scalar  and the resonance spectrum}

The most important goals of our lattice Higgs project are
to establish the emergence of the light scalar state with $0^{++}$ quantum numbers and the resonance spectrum perhaps far
separated from the light composite scalar.

\vskip 0.1in
\noindent{\bf The light scalar state}

The  ${\rm f_0}$ meson (in QCD terminology) has  ${\rm 0^{++}}$ quantum numbers and acts as the scalar state in the sextet model. 
Close to the conformal window, the ${\rm f_0}$ meson 
of the sextet model is not expected to be similar to its counterpart in QCD. If it turns out to be light, it could replace the elementary Higgs particle
and pose as the Higgs impostor.  
Two types of different ${\rm 0^{++}}$ operators, the fermionic one and the gluonic one (${\rm 0^{++}}$ glueball), are expected to mix. 
Such mixing was not included in the pilot study~\cite{Fodor:2014pqa} but becomes an important goal of our ongoing effort. 

%Implementation %1
%
\begin{figure}[hbt!]
	\begin{center}
		\scalebox{0.39}{\includegraphics{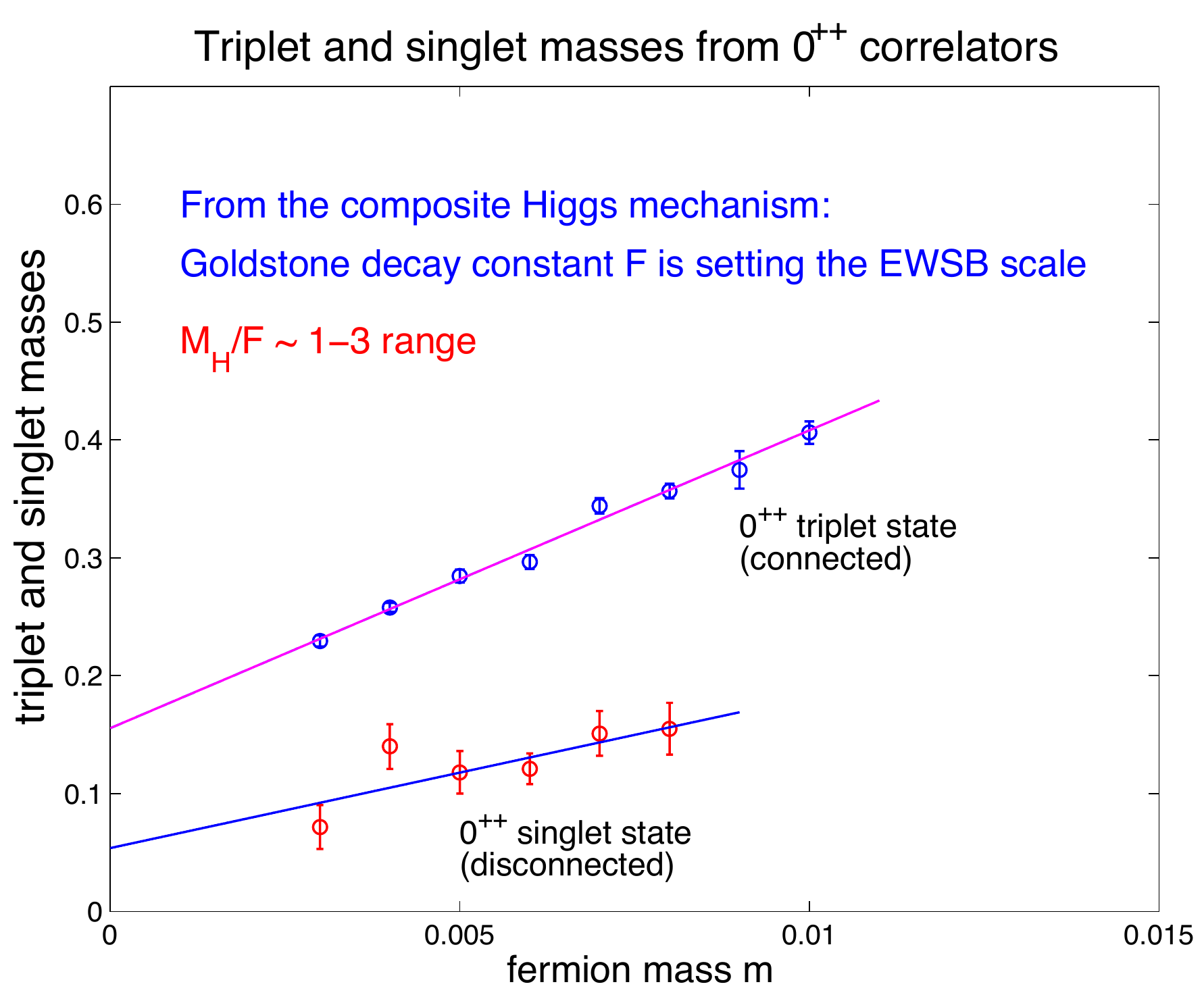}}
		\scalebox{0.495}{\includegraphics{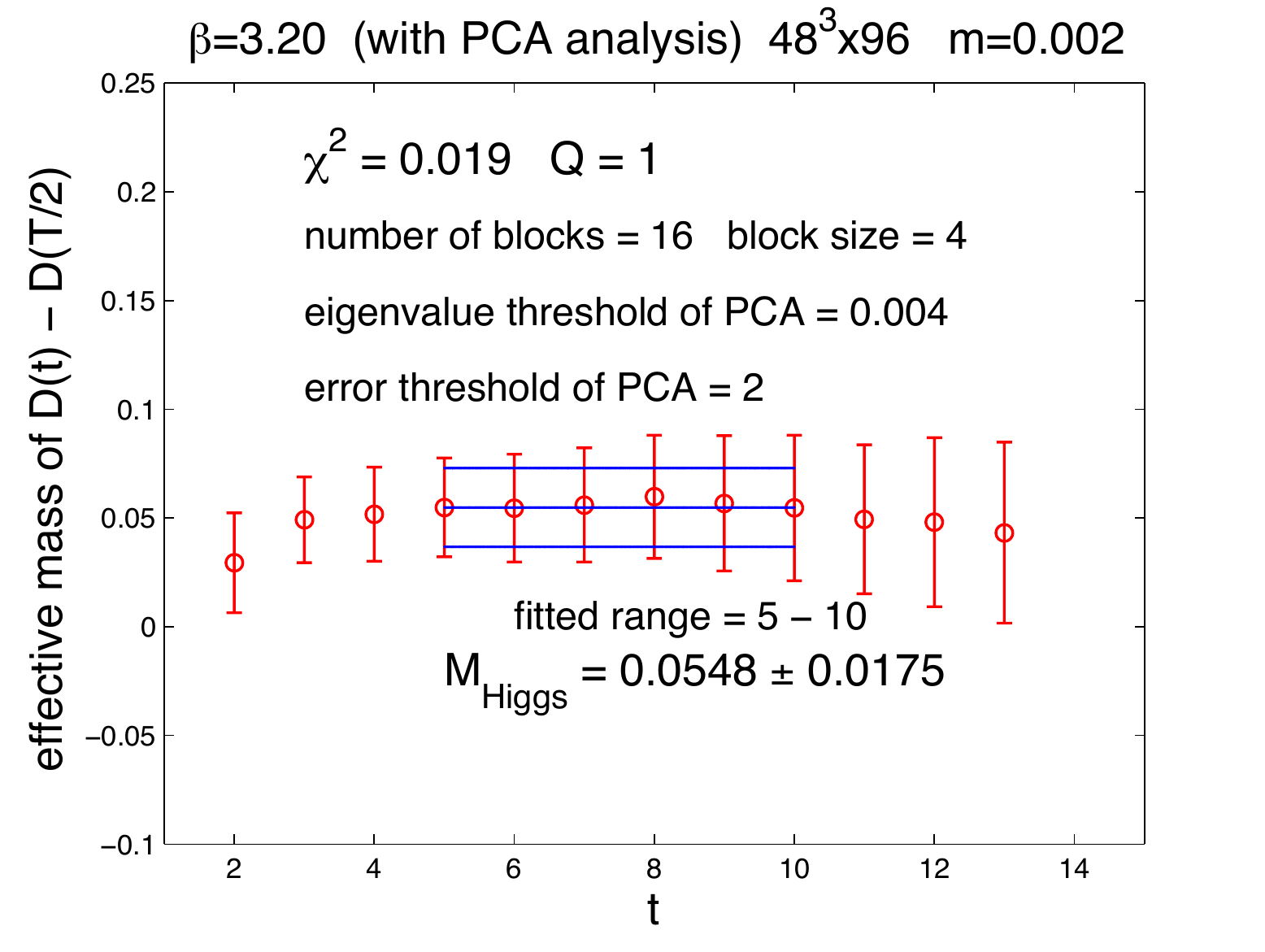}}
		\caption{\footnotesize The left panel shows earlier results on $32^3 \times 64$ lattice volumes at $\beta=3.20$~\cite{Fodor:2014pqa}. 	
		The right panel shows some representative new result on a large lattice volume using correlated fitting 
		and Principal Component Analysis (PCA).}\label{analysis}
%		\vskip -0.2in
	\end{center}
\end{figure}
A particular flavor-singlet correlator is needed to capture the ${\rm 0^{++}}$ scalar state with vacuum quantum numbers. It 
requires connected and disconnected diagrams of fermion loop propagators on the gauge configurations.
The connected diagram corresponds to the non-singlet correlator  ${\rm C_{\rm non-singlet}(t)}$. 
The correlator of the disconnected diagram is  ${\rm D(t)}$ at time separation ${\rm t}$. 
The ${\rm f_0}$ correlator ${\rm C_{\rm singlet}(t)}$ is defined as ${\rm C_{\rm singlet}(t) \equiv C_{\rm non-singlet}(t) + D(t)}$. 
The transfer matrix has the spectral decomposition of the ${\rm C_{\rm singlet}(t)}$ correlator in terms of the 
sum of all energy levels ${\rm E_i(0^{++}), i=0,1,2,...}$ and their  parity partners ${\rm E_j(0^{-+}), j=0,1,2,...}$ 
but at large time separation ${\rm t}$ the lowest states ${\rm E_0(0^{++}) }$ and ${\rm E_0(0^{-+})}$ dominate. 
They correspond to ${\rm m_{f_0}}$ and ${\rm m_{\eta_{\rm sc}}}$. 
The relevant non-singlet 
staggered correlator can be fitted well with non-oscillating $a_0$ contribution and oscillating $\pi_{\rm sc}$ contribution, with the 
non-Goldstone pion $\pi_{\rm sc}$ discussed in Section 3. 

We estimate the connected and disconnected diagrams with stochastic source vectors of fermion propagators.
To evaluate the disconnected diagram, we  need to calculate closed loops of quark propagators. 
We introduce $Z_2$ noise sources on the lattice where each source is defined on individual time-slice $t_0$ for color $a$. 
The scheme  can be viewed as a ``dilution'' scheme which is fully diluted in time and color and even/odd diluted in space.
The left panel of Figure~\ref{analysis} shows the earlier preliminary results from the pilot study 
on $32^3 \times 64$ lattice volumes at $\beta=3.20$~\cite{Fodor:2014pqa}.
From our new analysis a representative example of the scalar effective mass fit is shown at $\beta=3.20$ 
as the right panel of Figure~\ref{analysis}
on a large $48^3 \times 96$ lattice volume and probes the light scalar deeper toward the chiral limit at fermion mass ${\rm m=0.002}$.

Further work is needed on the light ${\rm f_0}$ scalar with ${\rm 0^{++}}$ quantum numbers.
% to clarify several important issues. 
The lower left panel of Figure~\ref{fig:condensate} indicates
large autocorrelation times in the topological history of the RHMC algorithm.
The effects of slowly changing topology on the fitted mass values of ${\rm f_0}$ are not sufficiently 
tested although sensitivity to the topological charge remains within the statistical accuracy of the runs. 
Fermion mass deformations of the low-lying ${\rm f_0}$ state and the Goldstone pion are expected to be entangled
which requires the modification of ${\rm \chi PT}$. Precise extrapolation 
to vanishing fermion mass in the chiral limit remains a challenging problem in the presence of the
light ${\rm f_0}$ state.

 \vskip 0.1in 
\noindent{\bf The emerging spectrum}

It is important to investigate the chiral limit of composite hadron states separated 
from the Goldstones and the light scalar by finite mass gaps. The baryon mass gap in the chiral limit provides further evidence
for ${\rm \chi SB}$ with preliminary results reported at this conference~\cite{Santanu:2015}.  
Resonance masses of parity partners provide important additional information with split parity masses in the chiral limit.
This is particularly important for consistency with ${\rm \chi SB}$ and for a first estimate of the S parameter 
when probing the model via Electroweak precision tests~\cite{Peskin:1991sw}.

\begin{figure}[ht!]
\begin{center}
\begin{tabular}{c}
\includegraphics[width=4.5in]{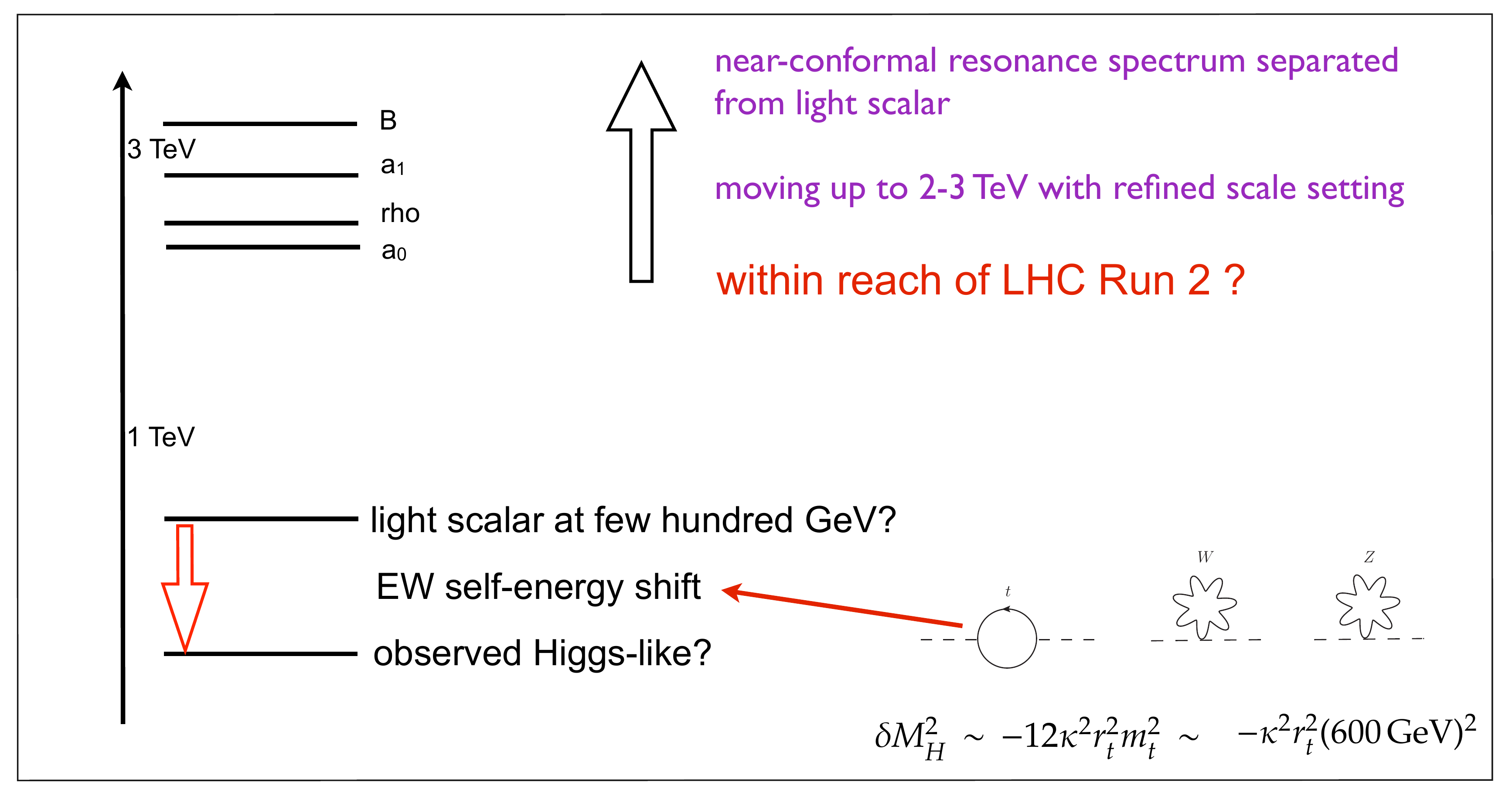}
\end{tabular}
\end{center}
\vskip -0.2in
\caption{\footnotesize  Schematic view of the emerging resonance spectrum.
The parameters ${\rm \kappa~and~r_t}$ are defined in ~\cite{Foadi:2012bb}.}
\label{fig:LHC}
\vskip -0.1in
\end{figure}

A remarkable spectrum is emerging which is sketched in Figure~\ref{fig:LHC} for illustration only.
Although with more work needed to confirm, the sextet model appears to be close to the conformal window and 
due to $\chi{\rm SB}$ exhibits
the right Goldstone spectrum for the minimal realization of the composite Higgs mechanism with a light scalar 
separated from the associated resonance spectrum in the 2 TeV region.
Chiral symmetry breaking and a very small beta function, perhaps slowly walking as hinted by preliminary results in Section 7,  are
not sufficient to guarantee a light dilaton-like state as the natural explanation for the emergence of the light scalar.
Consistent with our observations, a light Higgs-like scalar is still expected to emerge near the conformal window as a composite state
with $0^{++}$ quantum numbers, not necessarily with dilaton interpretation.
This scalar state has to be light but is not required to match exactly the observed ${\rm 126~GeV}$
mass. The light scalar
%dynamical Higgs mass ${\rm M^{0^{++}}_H}$ 
from composite strong dynamics gets lighter
from electroweak loop corrections,  dominated by the large negative mass shift from 
the top quark loop~\cite{Foadi:2012bb,Cacciapaglia:2014uja,DiChiara:2014uwa}.
%
%    

%\newpage
%\section{Mixed action on the gradient flow and the  ${\mathbf \epsilon}$ regime} 
 \section{Mixed action on the gradient flow and the epsilon regime}  

The alternative to safe extrapolation  from a regime of competing scalar and pion masses to the massless fermion limit
requires crossover to the epsilon regime of ${\rm \chi PT}$ at low enough scale ${\rm \lambda}$ where Goldstone dynamics begins 
to decouple from the scalar state.  This is difficult to do and requires significant resources. 
To control taste braking we cannot go to lattice spacings larger than the one set by ${\rm \beta=3.20}$. The value of ${\rm F\sim 0.025}$ at this 
lattice spacing requires large ${\rm 48^3\times 96}$ lattice volumes to control the ${\rm F\cdot L \geq 1}$ 
condition which is necessary for convergent expansion in all regimes of ${\rm \chi PT}$, including the epsilon regime. 
Even for our largest ${\rm V=48^3\times 96}$ and ${\rm V=40^3\times 80}$  lattice volumes control with
${\rm F\cdot L \sim 1-1.2}$ is just barely sufficient. 
For the lowest fermion mass  ${\rm m=0.002}$  the scaling variable ${\rm m\Sigma V\sim 100}$ is very large characterizing the p-regime  
of ${\rm \chi PT}$ we used earlier in the analysis. Reaching the epsilon regime would require an order of magnitude decrease in 
the scaling variable ${\rm m\Sigma V}$ 
which presents a considerable algorithmic challenge and requires substantial resources.  
Decreasing the fermion mass an order of magnitude
to ${\rm m=0.0002}$ would increase the cost,  scaling with ${\rm \sim 1/m}$ and calling for algorithmic improvements. 
\begin{figure}[htb!]
\begin{center}
\begin{tabular}{cc}
\includegraphics[height=5.5cm]{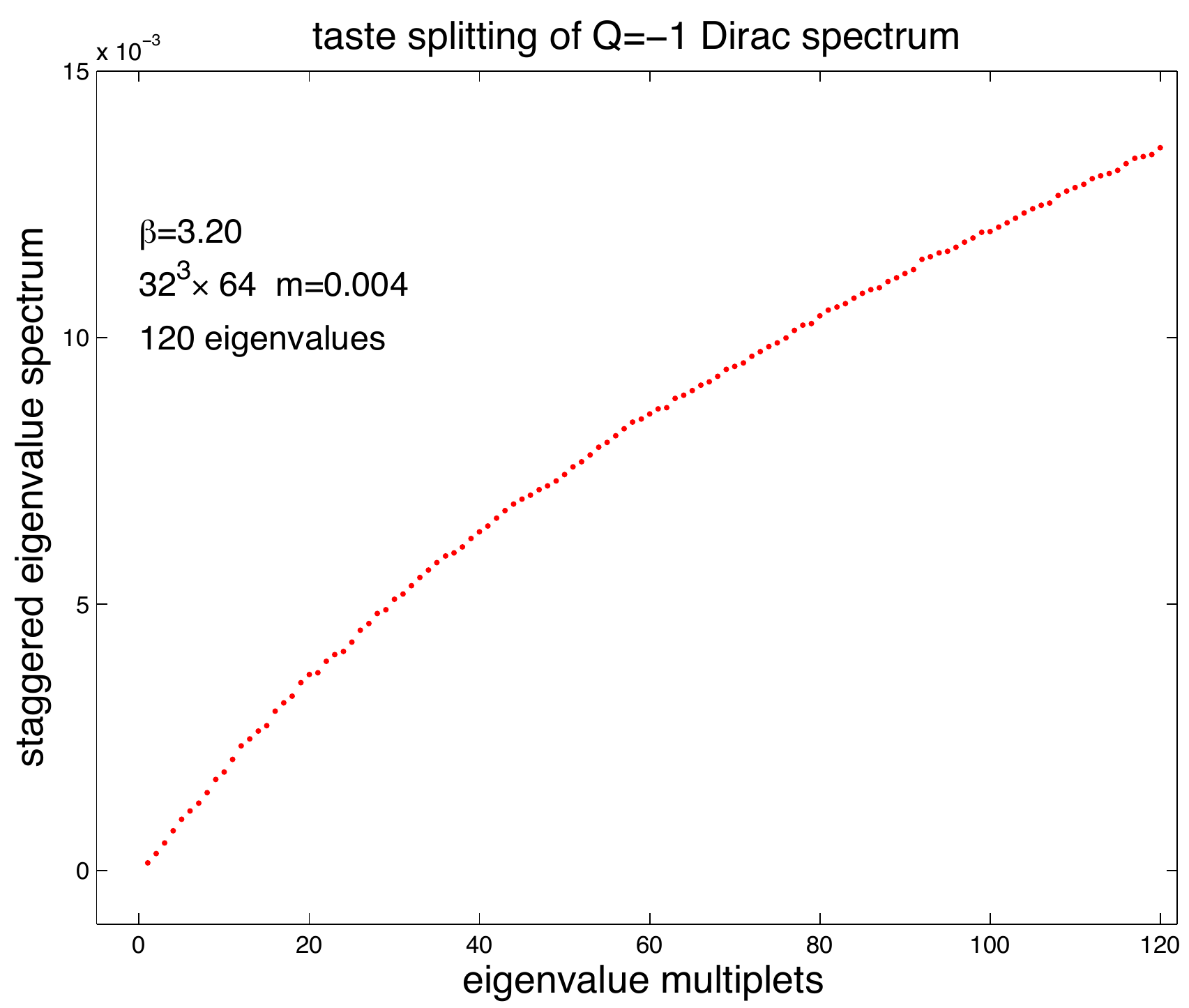}&
\includegraphics[height=5.5cm]{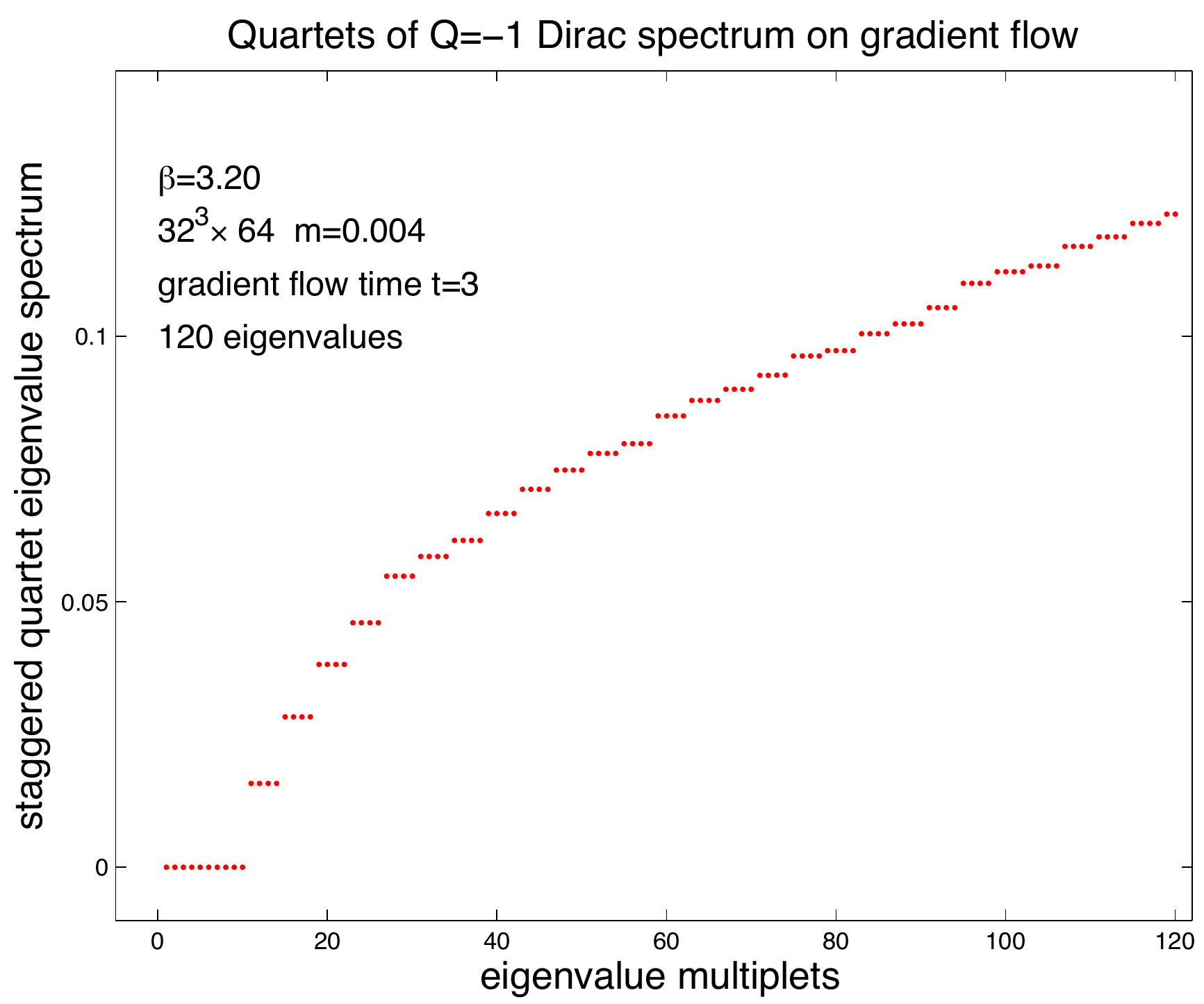}\\
\includegraphics[height=5.5cm]{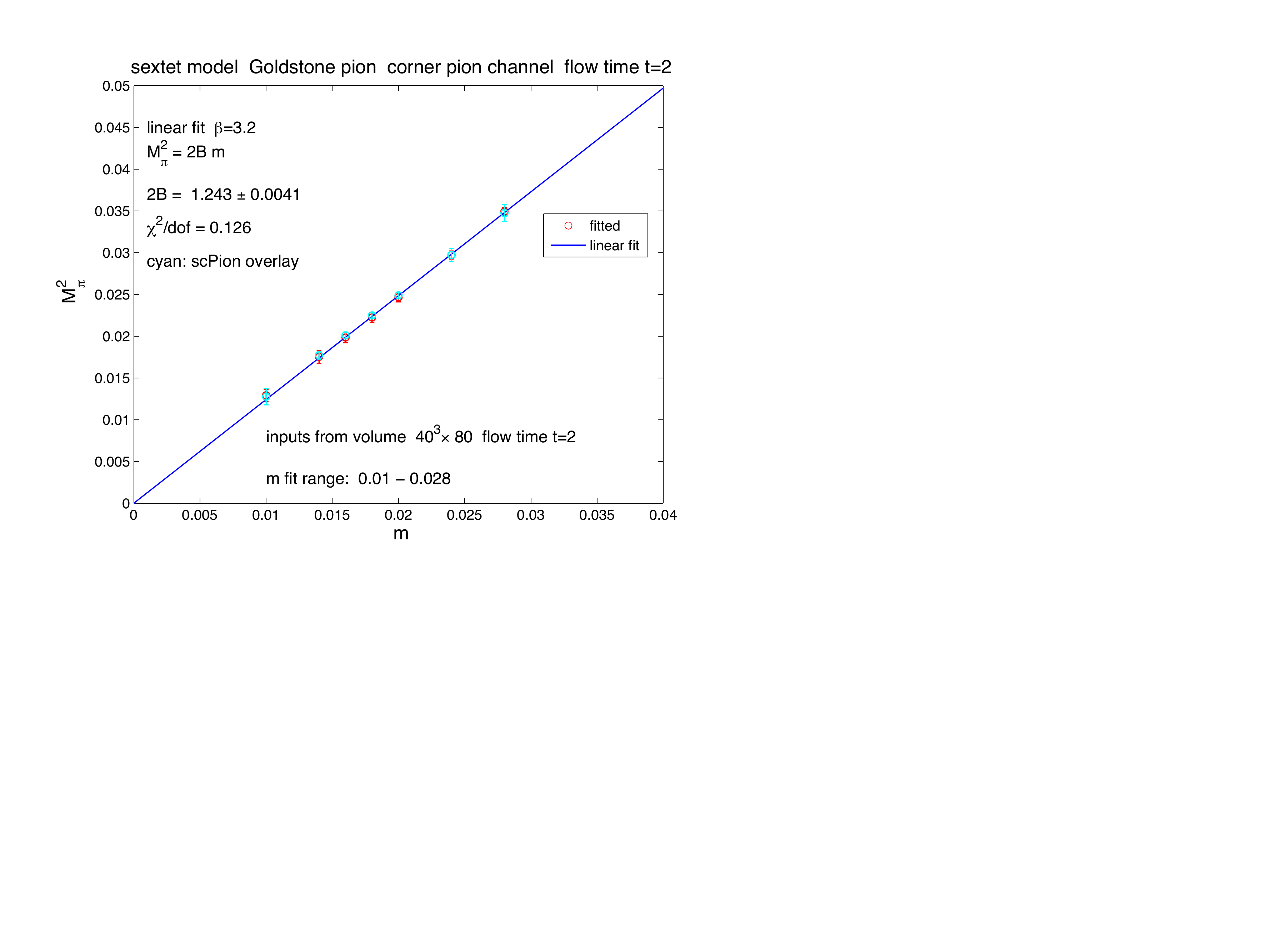}&
\includegraphics[height=5.5cm]{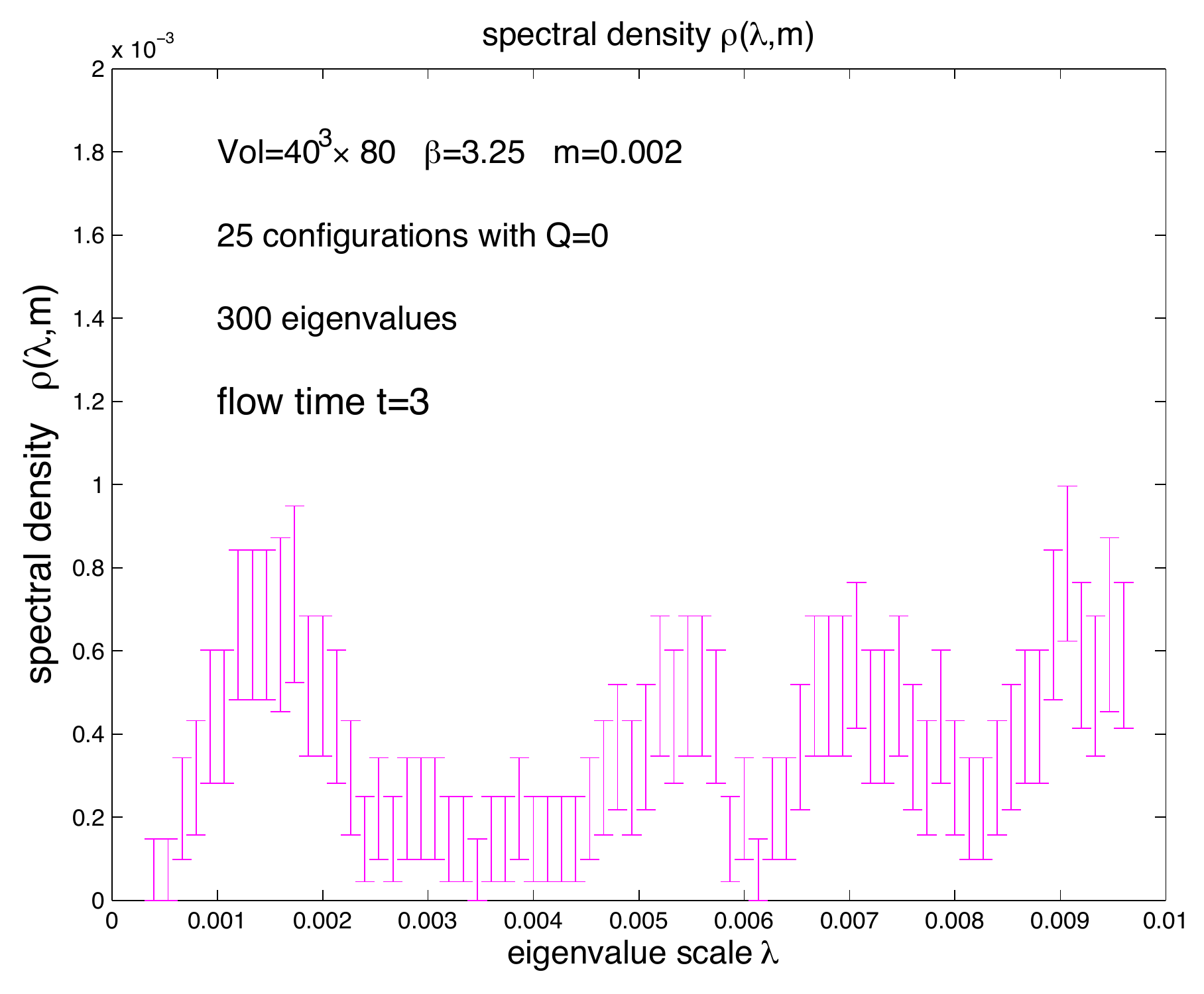}
\end{tabular}
\end{center}
\vskip -0.2in
\caption{\footnotesize  The upper right panel demonstrates the correct index theorem 
and the lower right panel displays the lowest eigenvalues in the epsilon regime.}
\label{fig:quartets}
%\vskip -0.2in
\end{figure}

We developed a promising new strategy to overcome the problem by performing ${\rm \chi PT}$ analysis in the crossover to the epsilon regime
with partial quenching and a related mixed action. 
We take the  p-regime gauge configurations of the lowest fermion masses on the largest lattice volumes
and analyze the fermion condensate and the Dirac spectrum after gradient flow times ${\rm t=2~or~3}$ with the valence fermion action 
where the original gauge link variables are replaced with the ones at flow time ${\rm t}$. This strategy 
can be viewed as a mixed action based analysis. The first encouraging results are shown in Figure~\ref{fig:quartets}.
The upper left panel shows the infrared part of the Dirac spectrum on the original gauge 
configurations with strong taste breaking evidenced by the absence of degenerate quartets. After gradient flow time ${\rm t=3}$
degenerate eigenvalue quartets emerge with the correct count of the topology dependent zero modes from the index theorem 
showing restored taste symmetry in the valence action. The lower left panel illustrates
the degeneracy of the Goldstone pion with non-Goldstone pions (scPion in the plot). The lower right panel shows that
the epsilon regime is reached with the scaling variable ${\rm \lambda\Sigma_{flow} V\sim 10}$ where the fermion mass is replaced by the scale 
of the Dirac spectrum (${\rm m\rightarrow\lambda)}$ and ${\rm \Sigma_{flow}}$, not RG invariant, is reduced by 
almost a factor of ten.
\vskip -0.8in

%\newpage
\section{The scale dependent renormalized coupling and beta function}

An important and independent consistency condition of the model would be provided
by matching the scale dependent renormalized coupling of the perturbative regime to the scale dependent 
coupling of the non-perturbative phase associated with  $\chi{\rm SB}$. 
We proposed a gauge coupling earlier
$g(\mu  = 1/L)$, running with the scale set by the finite volume~\cite{Fodor:2012td}
and defined on the gradient flow of the gauge field~\cite{Luscher:2010iy}.  
Since the gradient flow at flow time $t$ probes the gauge field on the scale $\sqrt{8t}$, the running coupling can be defined as a
function of $L$ in finite volume $V=L^4$ while holding $c=(8t)^{1/2}/L$  fixed:
%
%$\alpha_{c}(L)=\frac{4\pi}{3}\frac{\langle t^2E(t)\rangle}{1+\delta (c)}$.
${\rm \alpha_{c}(L)=4\pi\langle t^2E(t)\rangle /[3(1+\delta (c)]}$.
%
%
%\begin{equation}
%\alpha_{c}(L)=\frac{4\pi}{3}\frac{\langle t^2E(t)\rangle}{1+\delta (c)}\,.
%\label{eq:coupling}
%\end{equation}
%
\begin{figure}[h!]
	\begin{center}
		\scalebox{0.35}{\includegraphics{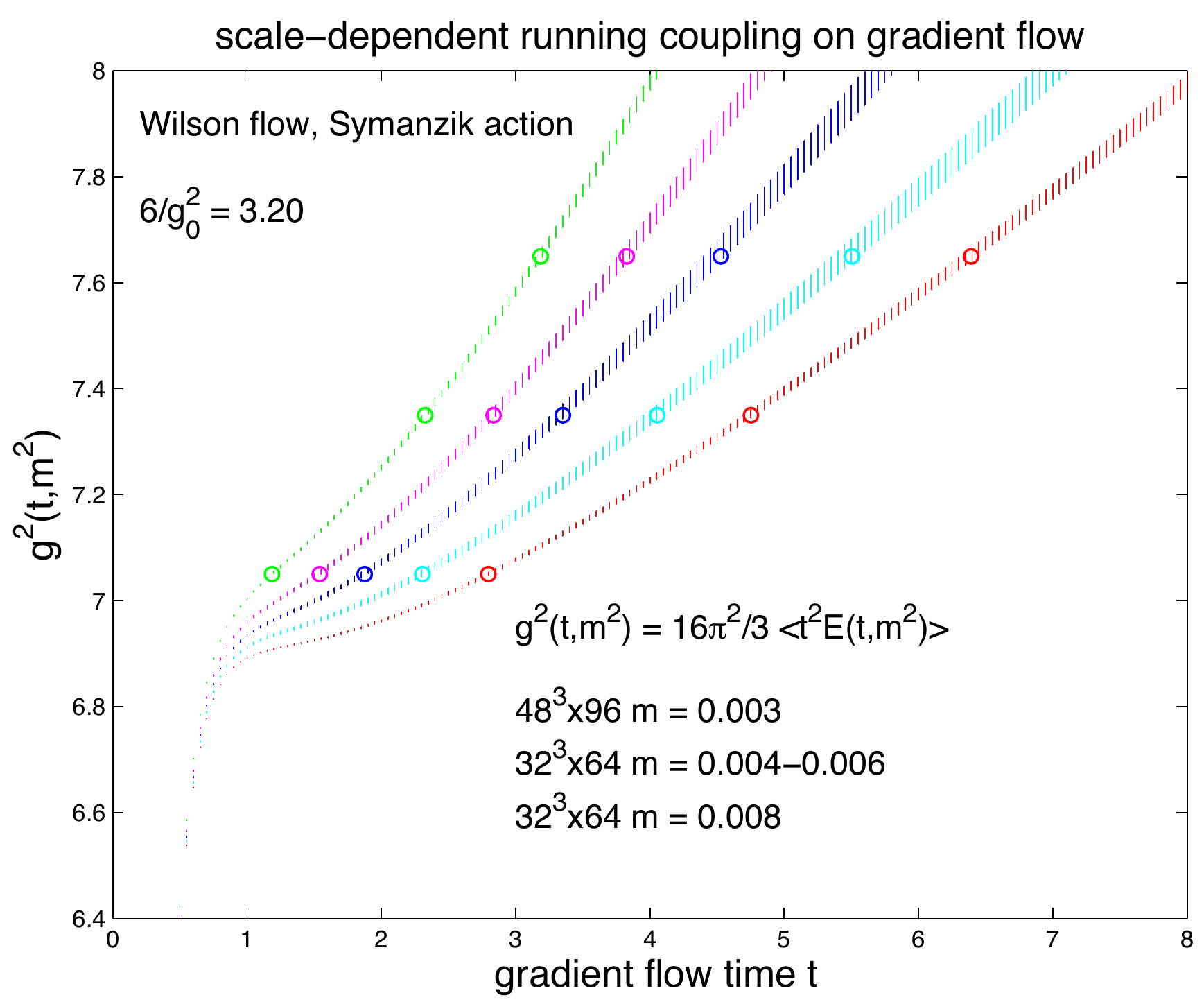}}
		\scalebox{0.35}{\includegraphics{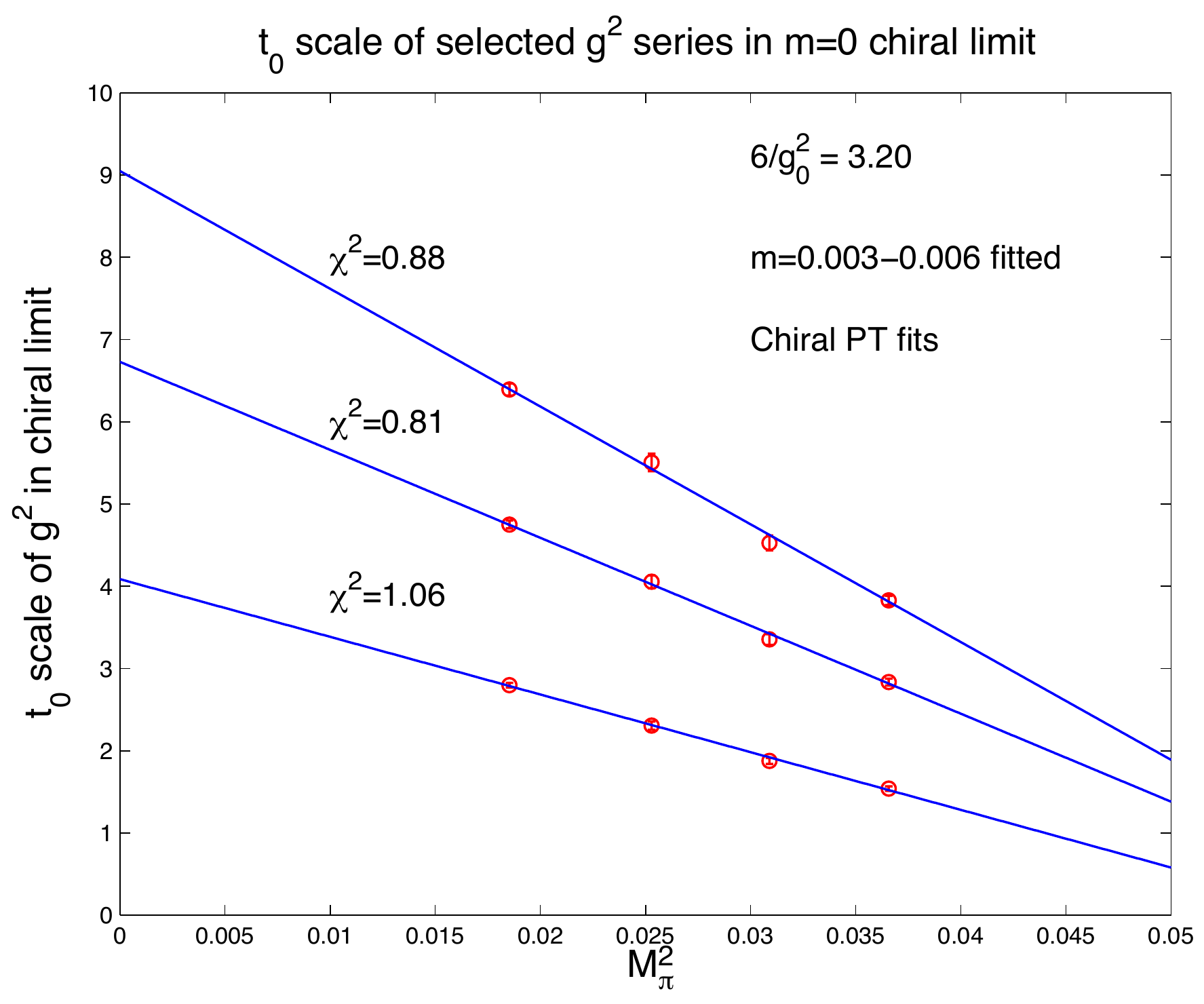}}\\
		\scalebox{0.35}{\includegraphics{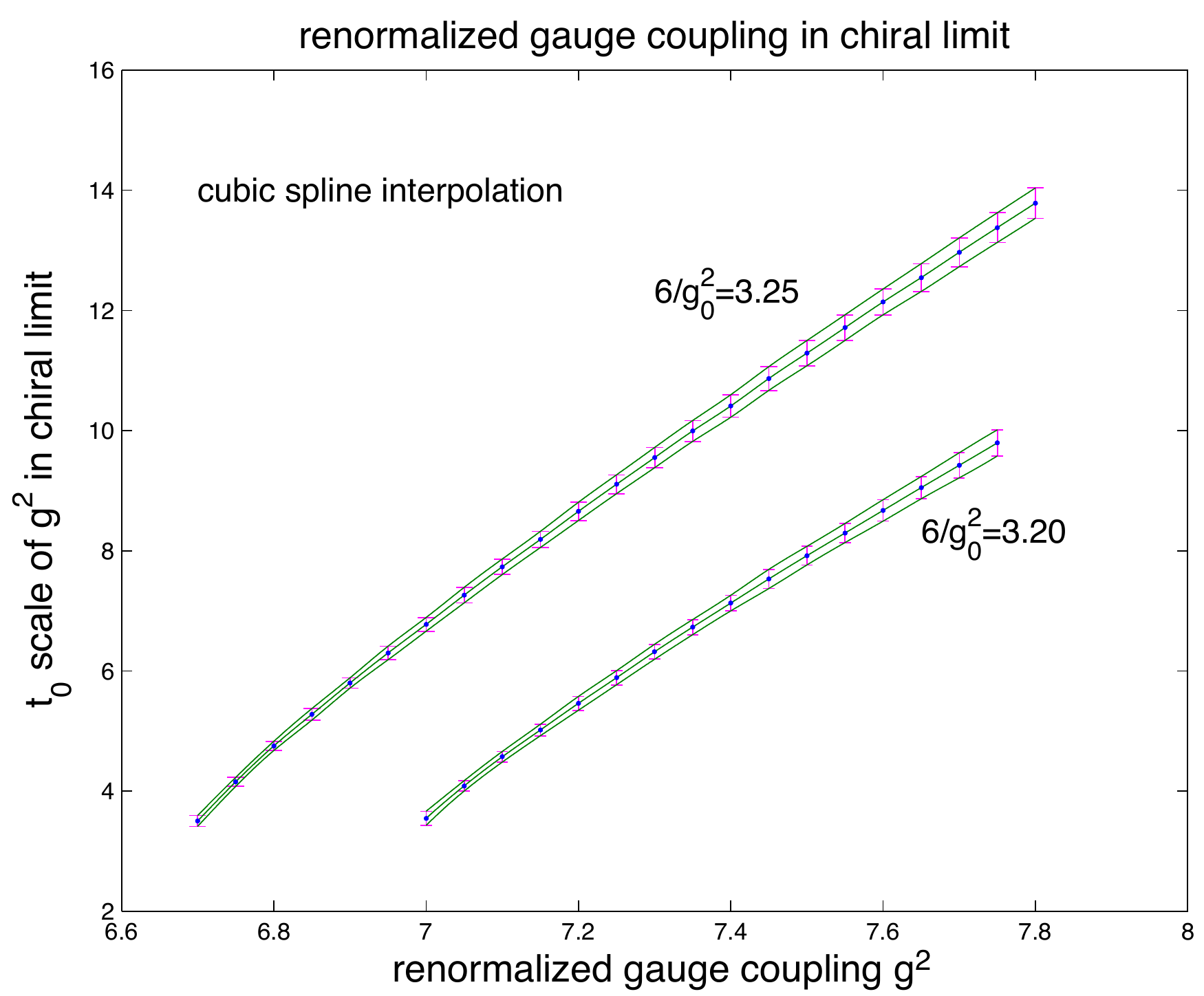}}
		\scalebox{0.35}{\includegraphics{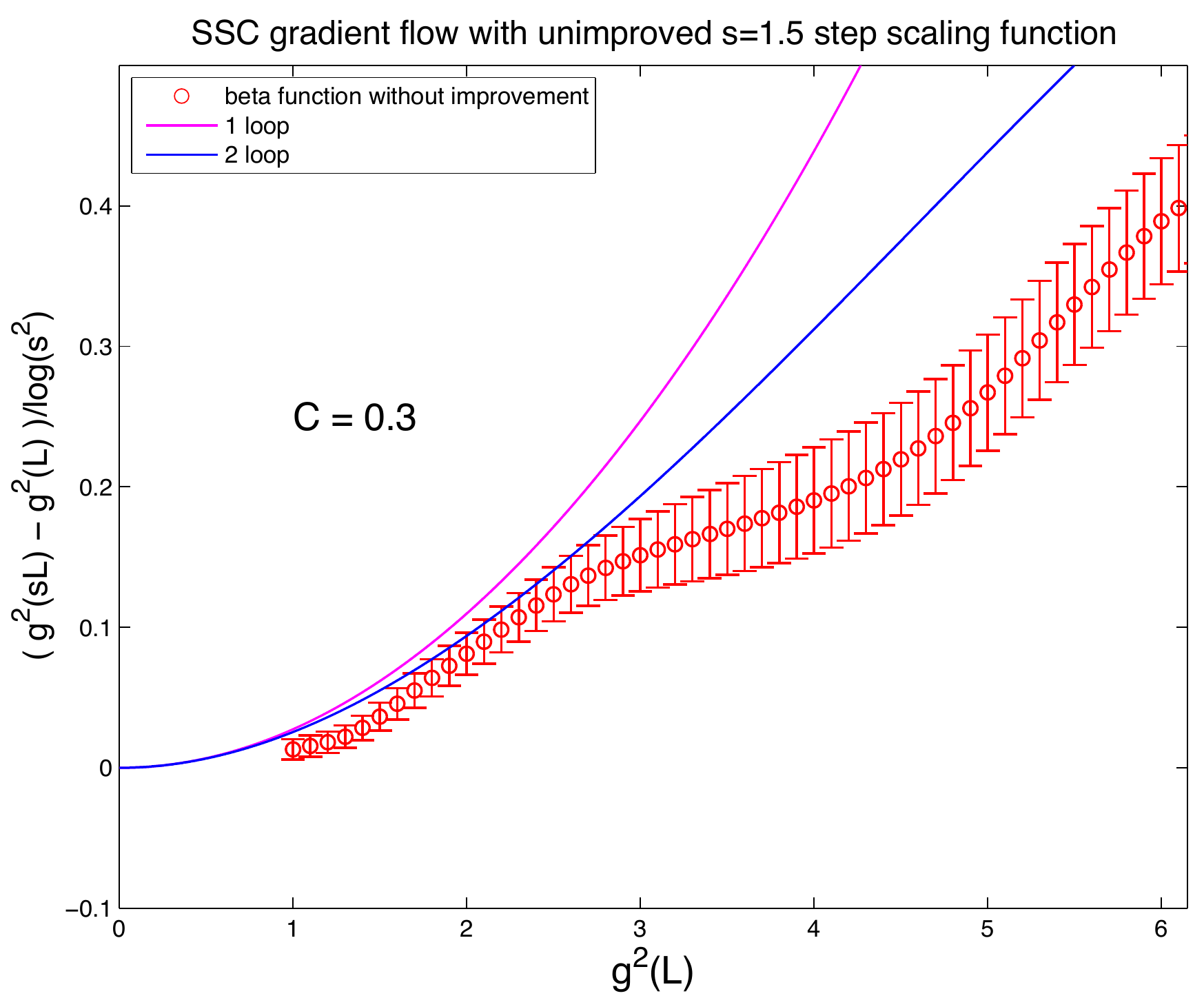}}
		\vskip -0.1in
		\caption{\footnotesize Two different schemes for scale dependent couplings 
		are illustrated with explanation in the text.}\label{fig:beta_function}
	\end{center}
	\vskip -0.15in
\end{figure}
This volume dependent coupling is 
particularly suitable to study the perturbative
regime and departures from it~\cite{Fodor:2012td}.
The measured renormalized couplings are very accurate and  the scheme defines a one-parameter family 
which can be adjusted for different goals~\cite{Fodor:2014cpa,Fodor:2014cxa}.
%with some of the latest results reported at the conference~\cite{Fodor:2014cxa}.
For illustration, preliminary post-conference results of the step beta function for the sextet model with two flavors 
are shown in the lower right panel of Figure~\ref{fig:beta_function}.

In the non-perturbative phase with ${\rm \chi SB}$ we are interested in a scale-dependent 
and volume independent renormalized coupling.
At fixed lattice size, bare coupling, and fermion mass m
we determine the appropriate flow time ${\rm  t(g^2, m)}$ to match any targeted 
flow-dependent renormalized coupling ${\rm g^2}$ calculated from ${\rm \langle t^2E(t)\rangle}$.
Assuming that the footprint of the operator  on the gradient flow is sufficiently small 
compared to the Compton wavelength of the pion for p-regime
analysis, the dependence of  ${\rm  t(g^2, m)}$ on ${\rm m}$ can be replaced by ${\rm  t(g^2, M^2_\pi)}$ in 
${\rm \chi PT}$ of pion dynamics with
linear dependence of  ${\rm  t(g^2, M^2_\pi)}$ on ${\rm M^2_\pi}$ in leading order~\cite{Bar:2013ora}.
Any residual finite volume dependence can be corrected in ${\rm \chi PT}$.

This strategy is illustrated by the step by step procedure in Figure~\ref{fig:beta_function}.
The upper left panel shows the determination of the flow time ${\rm  t(g^2, m)}$ of the targeted coupling ${\rm g^2}$
and the upper right panel is in surprisingly good agreement with the linear behavior in ${\rm M^2_\pi}$. 
At two different lattice spacings in the ${\rm m=0}$ chiral limit, the lower left panel shows the scale dependent renormalized 
coupling ${\rm g^2(t_0)}$ as a function of scale variation with  ${\rm t_0}$. A scale dependent and volume independent step 
beta function can be determined from this procedure.
A more comprehensive analysis of the data is part of our ongoing investigations including the extrapolation of the step function 
to vanishing lattice spacing and matching the two different scale dependent couplings of Figure~\ref{fig:beta_function}.

%Collecting more run parameter sets and a much more comprehensive analysis of the data is part of our ongoing
%investigations. The first analysis indicates that we should be able to extraplotae the step function to vanishing lattice
%spacing as the final goal. How to match this coupling to the lower right panel of the volume dependent coupling 
%remians an interesting challenge.

\vskip -0.5in
    
\section{Acknowledgement}
\vskip -.2cm
{\footnotesize We would like to thank Claude Bernard and Steve Sharpe for useful discussions 
on several aspects of the staggered ${\rm \chi PT}$ analysis.
We acknowledge support by the DOE under grant DE-SC0009919,
by the NSF under grants 0970137 and 1318220, by the EU Framework Programme 7 grant (F
P7/2007-2013)/ERC No 208740, by OTKA under the grant OTKA-NF-104034, and by the Deutsche
Forschungsgemeinschaft grant SFB-TR 55. Computational resources were provided by the DOE ALCC Award of
our collaboration  for the BG/Q Mira computational platform of
Argonne National Laboratory, by the BG/Q Juqueen platform of FZJ, by USQCD at Fermilab, by the University of Wuppertal, 
and by the Institute for Theoretical Physics,
E\"{o}tv\"{o}s University.
We are grateful to Szabolcs Borsanyi for helping us with his optimized code development for the BG/Q platform. 
We are also grateful to Kalman Szabo and Sandor Katz
for their code development building on Wuppertal gpu technology~\cite{Egri:2006zm}. }

\end{document}